\documentclass[dvips,arydshln]{./aa}
\usepackage{arydshln}
\usepackage{graphicx}
\usepackage{pslatex}
\usepackage{txfonts}
\usepackage{color}
\begin{document} 
%______________________________________________________________________________________
%______________________________________________________________________________________

\title {Resolving the clumpy circumstellar environment of the B[e] supergiant LHA 120-S 35 \thanks{Based on data acquired using 1) the du Pont Telescope at Las Campanas Observatory, Chile, under the programme CNTAC 2008-02 (PI: Barb\'a) 2) the MPG 2.2-m Telescope at La Silla Observatory, Chile, under the programme ID.: 094.A-9029(D) and under the agreement MPI-Observat\'orio Nacional/MCTIC, Prog. ID.: 096.A-9030(A), 3) the J. Sahade 2.15-m Telescope at Complejo Astron\'omico El Leoncito, operated under agreement between the Consejo Nacional de Investigaciones Cient\'ificas y T\'ecnicas de la Rep\'ublica Argentina and the National Universities of La Plata, C\'ordoba and San Juan, 4) the 8.1-m Telescope at Gemini South Observatory, which is operated by the Association of Universities for Research in Astronomy, Inc., under a cooperative agreement with the NSF on behalf of the Gemini partnership: the National Science Foundation (United States), the National Research Council (Canada), CONICYT (Chile), the Australian Research Council (Australia), Minist\'erio da Ci\v{e}ncia, Tecnologia e Inovac\~ao (Brazil) and Ministerio de Ciencia, Tecnolog\'ia e Innovaci\'on Productiva (Argentina), under the programme GS-2013B-Q-6 (PI: L. Cidale), 5) the Southern Astrophysical Research (SOAR) telescope, which is a joint project of the Minist\'{e}rio da Ci\^{e}ncia, Tecnologia, e Inova\c{c}\~{a}o (MCTI) da Rep\'{u}blica Federativa do Brasil, the U.S. National Optical Astronomy Observatory (NOAO), the University of North Carolina at Chapel Hill (UNC), and Michigan State University (MSU) and 6) the ESO Science Archive Facility.}}

%\subtitle{}

\author{A. F. Torres\inst{1,2}
\and L. S. Cidale\inst{1,2}
\and M. Kraus\inst{3,4} 
\and M. L. Arias\inst{1,2}
\and R. H. Barb\'a\inst{5}
\and G. Maravelias\inst{3,6}
\and M. Borges Fernandes\inst{7}}

\institute{Instituto de Astrof\'isica de La Plata (CCT La Plata - CONICET, UNLP), Paseo del Bosque S/N, La Plata, B1900FWA, Buenos Aires, Argentina\\ 
\email{atorres@fcaglp.unlp.edu.ar}\\
\and
Departamento de Espectroscop\'ia, Facultad de Ciencias Astron\'omicas y Geof\'isicas, Universidad Nacional de La Plata, Paseo del Bosque S/N, La Plata, B1900FWA, Buenos Aires, Argentina\\
\and
Astronomick\'y \'ustav, Akademie v\v{e}d \v{C}esk\'e republiky, Fri\v{c}ova 298, 251\,65 Ond\v{r}ejov, Czech Republic\\
\and
Tartu Observatory, T\~oravere, 61602 Tartumaa, Estonia\\
\and
Departamento de F\'isica y Astronom\'ia, Universidad de La Serena, Av. Cisternas 1200 Norte, La Serena, Chile\\
\and
Instituto de F\'{\i}sica y Astronom\'{\i}a, Universidad de Valpara\'{\i}so, Av. Gran Breta\~{n}a 1111, Casilla 5030, Valpara\'{\i}so, Chile\\
\and
Observat\'orio Nacional, Rua General Jos\'e Cristino 77, 20921-400 S\~ao Cristov\~ao, Rio de Janeiro, Brazil}

   \date{Received ...; accepted December 1, 2017.}
%______________________________________________________________________________________
%______________________________________________________________________________________

  \abstract
   { B[e] supergiants are massive post-main-sequence stars, surrounded by a complex circumstellar environment where molecules and dust can survive. The shape in which the material is distributed around these objects and its dynamics as well as the mechanisms that give rise to these structures are not well understood.}
   { The aim of this work is to deepen our knowledge of the structure and kinematics of the circumstellar disc of the B[e] supergiant \object{LHA\,120-S\,35}.}
   { High-resolution optical spectra were obtained in three different years. Forbidden emission lines, that contribute to trace the disc at different distances from the star, are modelled in order to determine the kinematical properties of their line-forming regions, assuming Keplerian rotation. In addition, we used low-resolution near-infrared spectra to explore the variability of molecular emission.}
   {\object{LHA\,120-S\,35} displays an evident spectral variability in both optical and infrared regions. The P-Cygni line profiles of \ion{H}{i}, as well as those of \ion{Fe}{ii} and \ion{O}{i}, suggest the presence of a strong bipolar clumped wind. We distinguish density enhancements in the P-Cygni absorption component of the first Balmer lines, which show variations in both velocity and strength. The P-Cygni profile emission component is double-peaked, indicating the presence of a rotating circumstellar disc surrounding the star. We also observe line-profile variations in the permitted and forbidden features of \ion{Fe}{ii} and \ion{O}{i}. In the infrared, we detect variations in the intensity of the \ion{H}{i} emission lines as well as in the emission of the CO band-heads. Moreover, we find that the profiles of each [\ion{Ca}{ii}] and [\ion{O}{i}] emission lines contain contributions from spatially different (complete or partial) rings. Globally, we find evidence of detached multi-ring structures, revealing density variations along the disc. We identify an inner ring, with sharp edge, where [\ion{Ca}{ii}] and [\ion{O}{i}] lines share their forming region with the CO molecular bands. The outermost regions show a complex structure, outlined by fragmented clumps or partial-ring features of \ion{Ca}{ii} and \ion{O}{i}. Additionally, we observe variations in the profiles of the only visible absorption features, the \ion{He}{i} lines.}
   {We  suggest that \object{LHA\,120-S\,35} has passed through the red-supergiant (RSG) phase and evolves back bluewards in the Hertzsprung-Russell diagram. In this scenario, the formation of the complex circumstellar structure could be the result of the wind-wind interactions of the post-RSG wind with the previously ejected material from the RSG. The accumulation of material in the circumstellar environment could be attributed to enhanced mass-loss, probably triggered by stellar pulsations. However, the presence of a binary companion can not be excluded. Finally, we find that \object{LHA\,120-S\,35} is the third B[e] supergiant belonging to a young stellar cluster.}


   \maketitle

%______________________________________________________________________________________
%______________________________________________________________________________________
% FIGURE -1-----------------------------
\begin{figure*}[!t]   
  \begin{centering}
  \includegraphics[angle=270,width=0.9\textwidth]{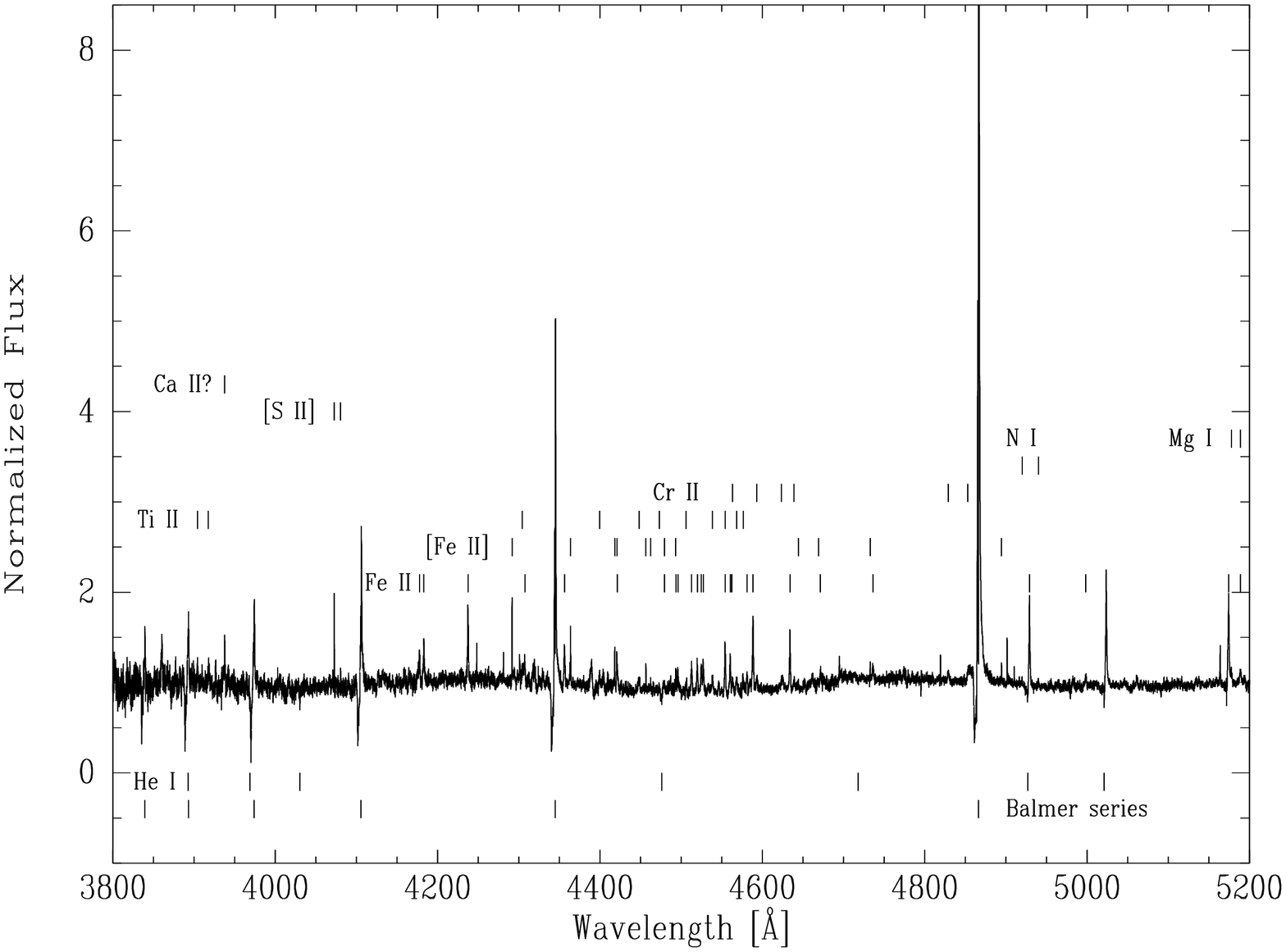}
  \caption{FEROS spectrum from 2014 covering 3800--5200 \AA\,. Main emission lines are indicated. \ion{He}{i} lines are in absorption.}
  \label{Figure-3800-5200-lines}
  \end{centering}
\end{figure*}
%--------------------------------------
% FIGURE 0-----------------------------
\begin{figure*}[!t]   
  \begin{centering}
  \includegraphics[angle=270,width=0.9\textwidth]{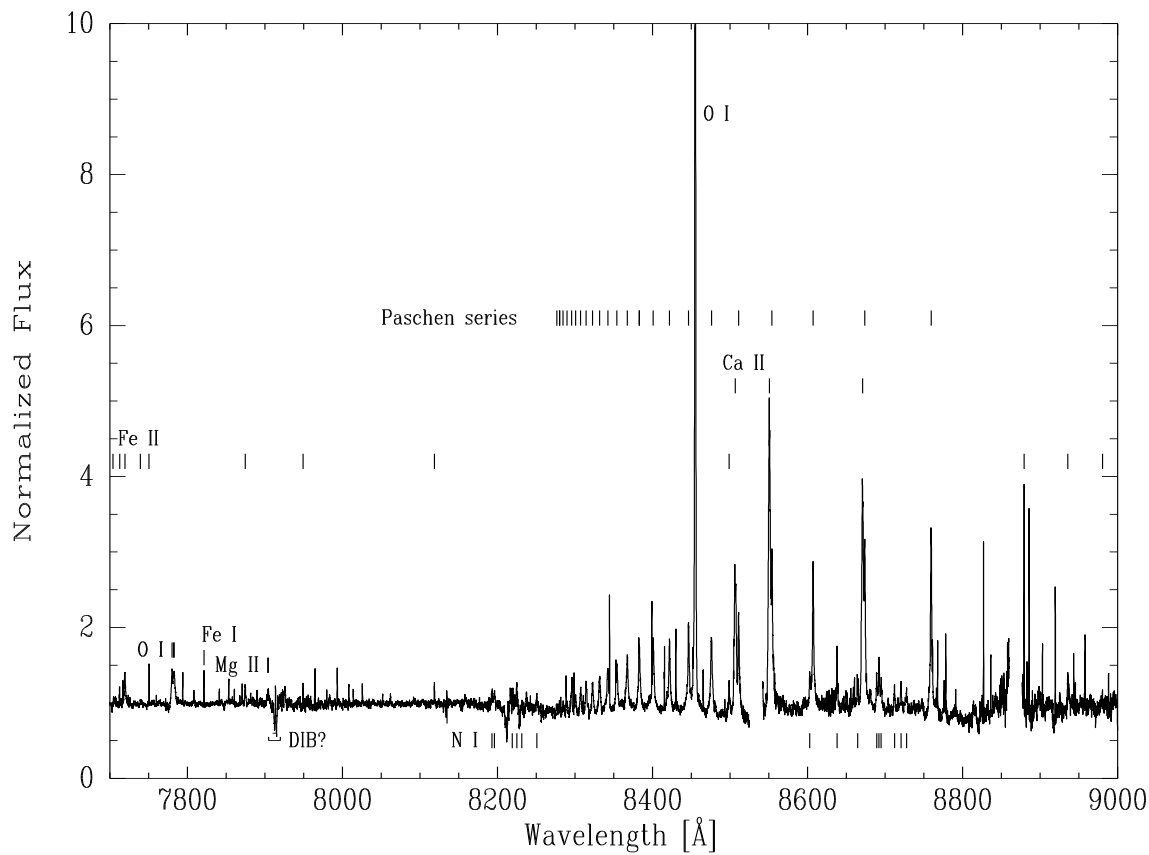}
  \caption{FEROS spectrum from 2014 covering 7700--9000 \AA\,. Main emission lines are indicated. Two absorption features are visible, one could be attributed to a diffuse interstellar band (DIB) while the other remains unidentified.}
  \label{Figure-7700-9000-lines}
  \end{centering}
\end{figure*}
%--------------------------------------
% FIGURE 1-----------------------------
\begin{figure*}[!t]   
  \begin{centering}
 \includegraphics[angle=270,width=0.9\textwidth]{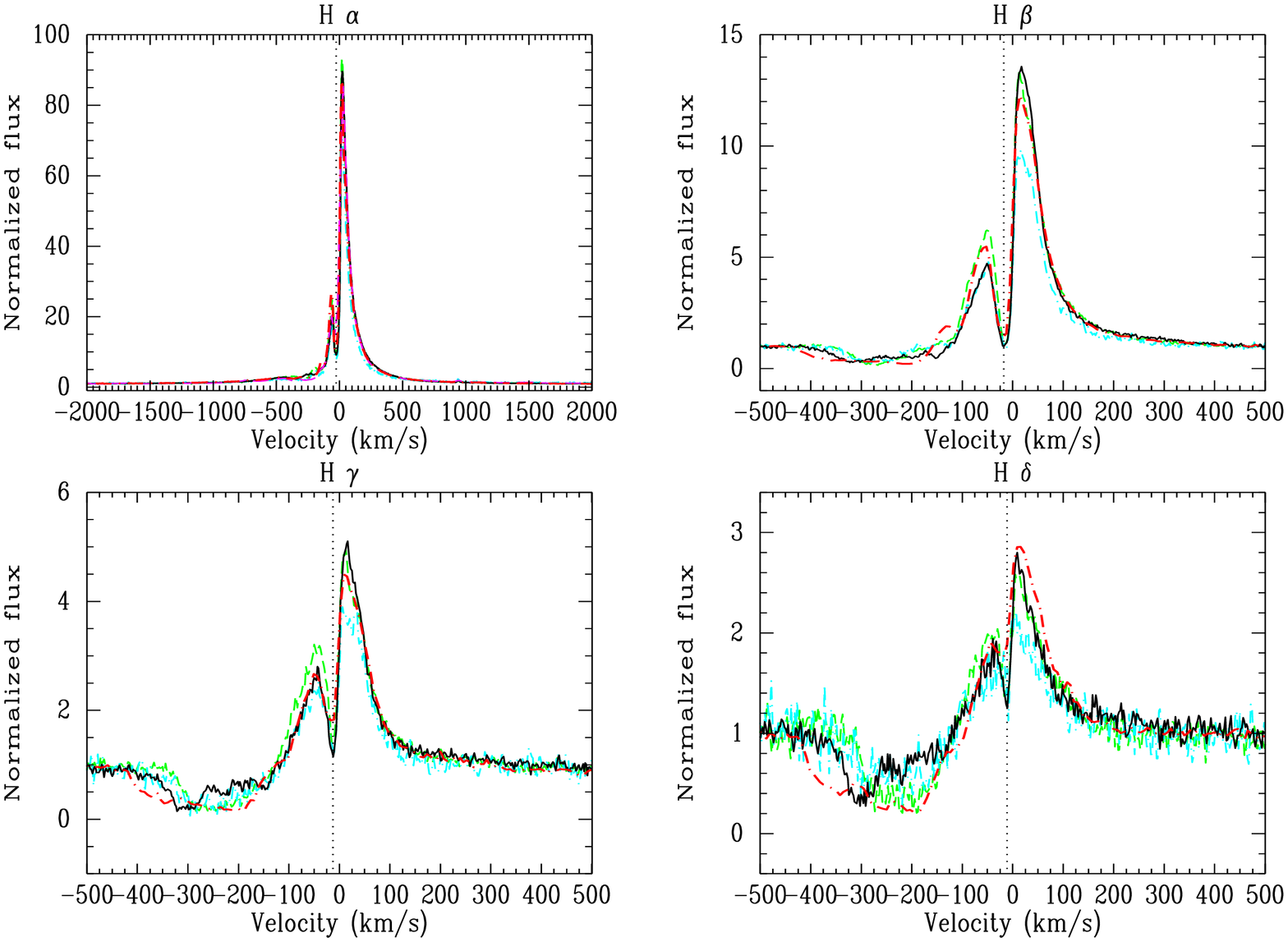}
  \caption{Very complex P-Cygni profiles of the first Balmer lines seen in the spectrum of \object{LHA\,120-S\,35} on a velocity scale relative to $V_{\rm{sys}}$ = 308 km s$^{-1}$. The spectrum in red dash-dotted line was taken at LCO in 2008, while the ones in black solid, cyan dashed and green dash-dotted lines correspond to the observations acquired at ESO in 2014, Nov 2015 and Dec 2015, respectively. In the H$\alpha$ plot, we also included the spectrum acquired at CASLEO in 2012 in magenta dashed line. This colour coding will be used in the rest of the plots. The central absorption of the double-peaked emission component of the H$\alpha$, H$\beta$, H$\gamma$ and H$\delta$ lines is at $\sim$ -25 km s$^{-1}$, -17 km s$^{-1}$, -13 km s$^{-1}$ and -11 km s$^{-1}$, respectively (indicated by a vertical dotted line).}
  \label{Figure-balmer-lines}
  \end{centering}
\end{figure*}
%-------------------------------------
%______________________________________________________________________________________
%______________________________________________________________________________________

\section{Introduction}

The importance of studying massive stars in astrophysics is evident as they play an important role in the galactic evolution. These stars may show strong mass-loss rates, which affect their own evolution, as well as the chemistry and dynamics of their surrounding medium. Sometimes the released material is accumulated forming circumstellar envelopes with particular physical conditions that lead to the development of the B[e] phenomenon. The mean properties of this phenomenon are the noticeable emission lines in the optical spectra of B-type stars, typically low-excitation permitted and forbidden transitions in neutral or low-ionized metals, and a strong infrared excess due to hot circumstellar gas and dust \citep{con97}. The B[e] phenomenon appears in many different stages of evolution, ranging from the pre-main-sequence to the planetary nebula or supergiant stages \citep{lam98}.

Our interest is focussed on B[e] supergiant stars (B[e]SGs), that constitute a homogeneous group among evolved stars that show the B[e] phenomenon. The study of this particular group can shed light on a short post-main-sequence evolutionary phase characterized by strong mass-loss rates. The ejected material of B[e]SGs shows two different regions: a fast polar wind structure and either a Keplerian or slow expanding equatorial disc or ring. The disc configuration has been confirmed by polarimetric and interferometric observations \citep{mag92,oud99,mel01,mag06,dom07,mil11,cid12}. These discs provide ideal conditions for molecule formation and dust condensation \citep{kra00,sta01}. 

The number of B[e]SGs known nowadays is very small: two in M31, 15 in the Magellanic Clouds, and around 15 candidates in our Galaxy \citep{kra09, Lev14, Kra14}, and although some of them are well studied many questions still remain unclear. So, an exhaustive research on a particular object can give us significant information about the physical properties of the circumstellar medium that could help in a near future to infer global properties of this particular group. To achieve this goal, we chose \object{LHA\,120-S\,35}, a poorly studied star of the Large Magellanic Cloud (LMC).

\object{LHA\,120-S\,35} was identified as S\,35 by \citet{hen56} in his catalogue of H$\alpha$ emission stars and nebulae. \citet{gum95} classified this object as a B[e] supergiant and derived the following stellar parameters: $T_{\mathrm {eff}}$ = 22\,000 K, $\log\,g$ = 3.0, $E(B-V)$ = 0.06, $L_*$ = 1.6 $\times$ 10$^5$ $L_{\odot}$, $R_*$ = 28 $R_{\odot}$, and a ZAMS mass of $M_* \sim$ 22 $M_{\odot}$, based on the fitting of theoretical fluxes to the observed stellar continuum from the ultraviolet to the infrared. They also analysed a high-resolution optical spectrum and reported the presence of permitted and forbidden emission lines (of Fe\,{\sc ii}, Ti\,{\sc ii}, Cr\,{\sc ii}, [Fe\,{\sc ii}], and [S\,{\sc ii}]), as well as strong Balmer lies with complex P Cygni-type profiles. From archival observations of the Far Ultraviolet Spectrographic Explorer (FUSE), \citet{pen09} estimated the projected rotational velocity, $v \sin\,i$ = 159 km s$^{-1}$. Bonanos et al. (\citeyear{bon09}) presented its spectral energy distribution from 0.3 $\mu$m to 24 $\mu$m, showing infrared features that confirmed the presence of dust. Infrared emission at 70 $\mu$m was also detected. Recently, this star was included in the $K$-band SINFONI (Spectrograph for INtegral Field Observations in the Near Infrared) survey presented by \citet{oks13}, who reported the first detection of the CO band head emission at 2.3 $\mu$m. Assuming that CO molecules are located in a narrow ring around the star, they found from model fittings to the observed CO band head emission a CO column number density of $N_{\mathrm {CO}} \sim$ 2 $\times$ 10$^{21}$ cm$^{-2}$ and a temperature of $T_{\mathrm {CO}} \sim$ 3\,000 K. This temperature is much lower than the CO dissociation temperature (5\,000 K) and  suggests that the material may be located in a detached disc structure \citep{lie10}. \citet{oks13} also noticed the presence of \element[ ][13]{CO} lines and derived a \element[ ][12]{C}/\element[ ][13]{C} ratio of around ten and interpreted that \object{LHA\,120-S\,35} is close to entering the red supergiant phase or perhaps evolving bluewards after this phase.

In this work we report on the unpublished spectral appearance of \object{LHA\,120-S\,35} longwards to 5200 \AA\, which provides valuable additional information to the understanding of this peculiar object. The paper is organized as follows: in Section \ref{sec:observations}, we present our optical and infrared observations; in Section \ref{sec:results}, we present an analysis of the main spectral features; in Section \ref{sec:kinematics}, we model the kinematics of the circumstellar material. In Section \ref{sec:discussion} we discuss our results in the context of possible scenarios and present our conclusions.

%______________________________________________________________________________________

\section{Observations}\label{sec:observations}

\subsection{Optical spectra}\label{sec:optical}

Two high-resolution optical spectra of LHA 120-S 35 ($R \sim$ 45\,000) were obtained on 15 November 2008, with the echelle spectrograph at the 2.5-m du Pont Telescope at Las Campanas Observatory (LCO) in Chile. We selected the following configuration: a  $1 \times 4$ arcsec slit and a Tek5 2k $\times$ 2k CCD detector, with a pixel size of 24 $\mu$m, and a 2 $\times$ 2-pixel binning. The spectral coverage ranges from 3600 \AA\, to 9200 \AA. The exposure time was 2400 s for each spectrum, obtaining a signal-to-noise ratio (S/N) per pixel in the 5500 \AA\, region of $\sim$ 60. Data were reduced using standard IRAF\footnote{IRAF is distributed by the National Optical Astronomy Observatories, which are operated by the Association of Universities for Research in Astronomy, Inc., under cooperative agreement with the National Science Foundation.} tasks. Spectra were bias subtracted, flat-field normalized, and wavelength calibrated.

Four high-resolution optical spectra of LHA 120-S 35 were acquired on 3 December 2014 and 4 December 2015 (two spectra per night) with the Fiber-fed Extended Range Optical Spectrograph (FEROS), attached to the MPG/ESO 2.2-m telescope at La Silla Observatory, Chile. FEROS is a bench-mounted echelle spectrograph, which provides data with a resolving power $R \sim$ 48\,000 and a spectral coverage from 3600 \AA\, to 9200 \AA. An EEV 2k $\times$ 4k CCD detector with a pixel size of 15 $\mu$m was used. The exposure time of each spectrum was 1300 s. The spectra were added to achieve a better S/N ratio. The final S/N was $\sim$ 30 per pixel in the 5500 \AA\, region. The reduction process was performed using the FEROS standard on-line reduction pipeline. 

%______________________________________________________________________________________

% FIGURE 2-----------------------------
\begin{figure}    
  \begin{centering}
  \includegraphics[angle=270,width=0.5\textwidth]{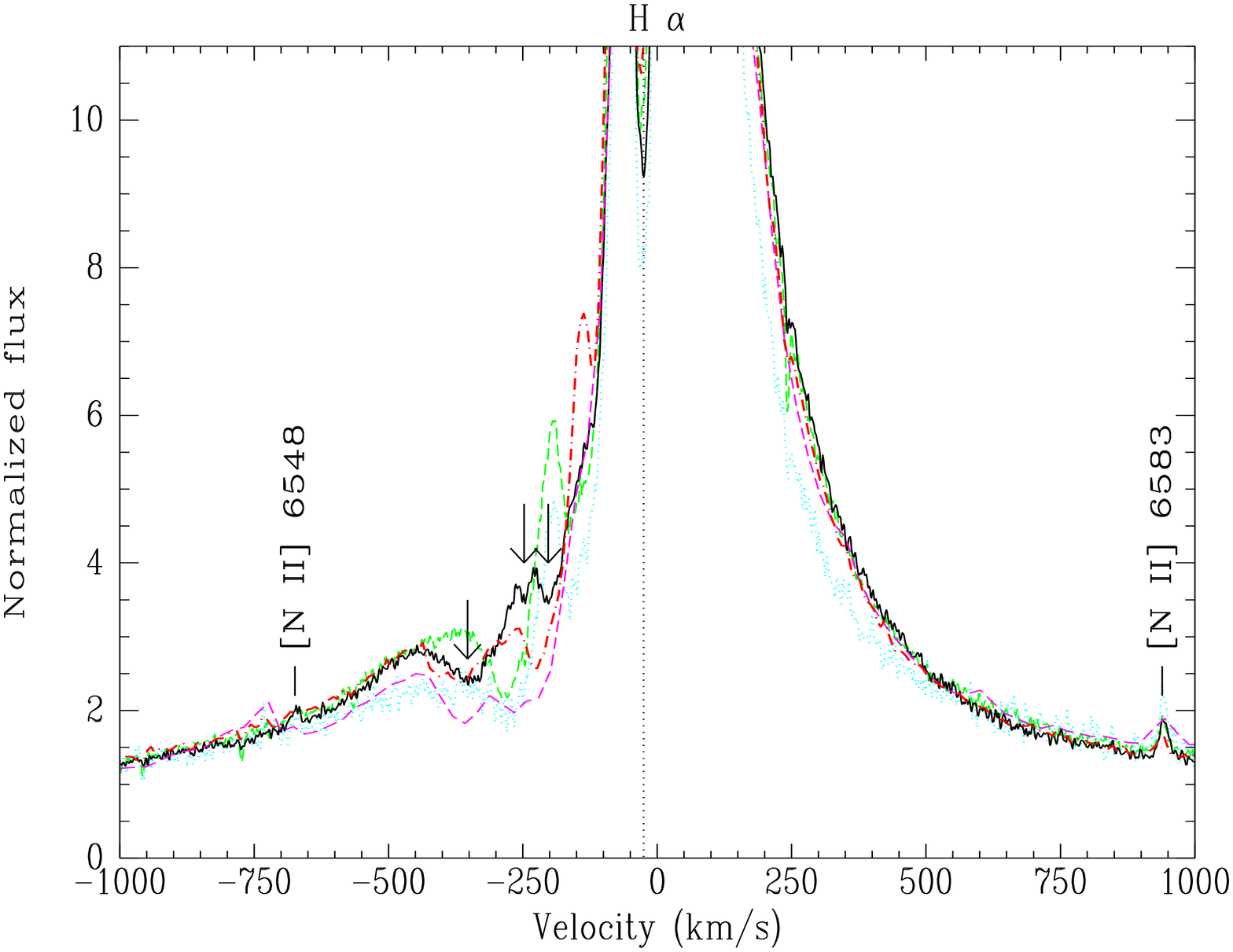}
  \caption{Zoom of the very complex profile of the H$\alpha$ line of \object{LHA\,120-S\,35}. A deep central absorption component at $\sim$ -25 km s$^{-1}$ is indicated by a vertical dotted line. Narrow absorption components are marked with arrows on the 2014 FEROS spectrum. The [\ion{N}{ii}] $\lambda\lambda$\,6548, 6583 lines are labelled. The same colour coding used in Fig. \ref{Figure-balmer-lines} is applied for Figs. \ref{Figure-halpha-line}-\ref{Figure-caii-lines}.}
  \label{Figure-halpha-line}
  \end{centering}
\end{figure}
%-------------------------------------
% FIGURE 3-----------------------------
\begin{figure}
  \begin{centering}
  \includegraphics[angle=270,width=0.5\textwidth]{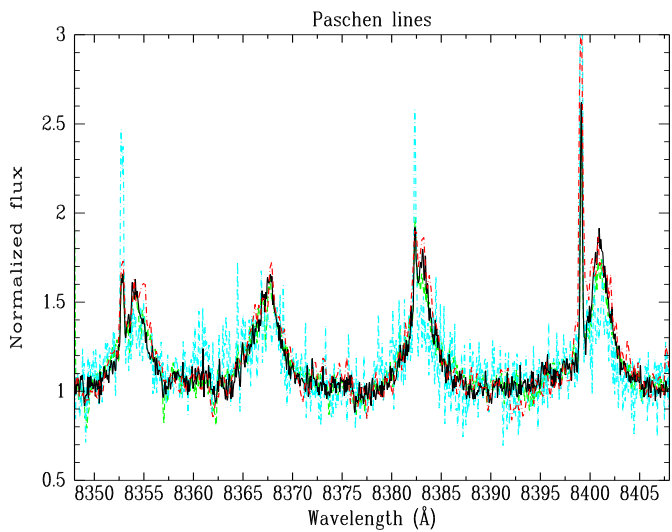}
  \caption{Example of the Paschen series lines (Pa23, Pa22, Pa21 and Pa20 lines, from left to right) of \object{LHA\,120-S\,35}. The sharp and narrow components seen on the left of the wide profiles could be related to the LMC background diffuse emission.}
  \label{Figure-paschen-lines}   
  \end{centering}
\end{figure}
%-------------------------------------
%______________________________________________________________________________________

%______________________________________________________________________________________

% FIGURE 4-----------------------------
\begin{figure}[!t]
  \begin{centering}
  \includegraphics[angle=270,width=0.5\textwidth]{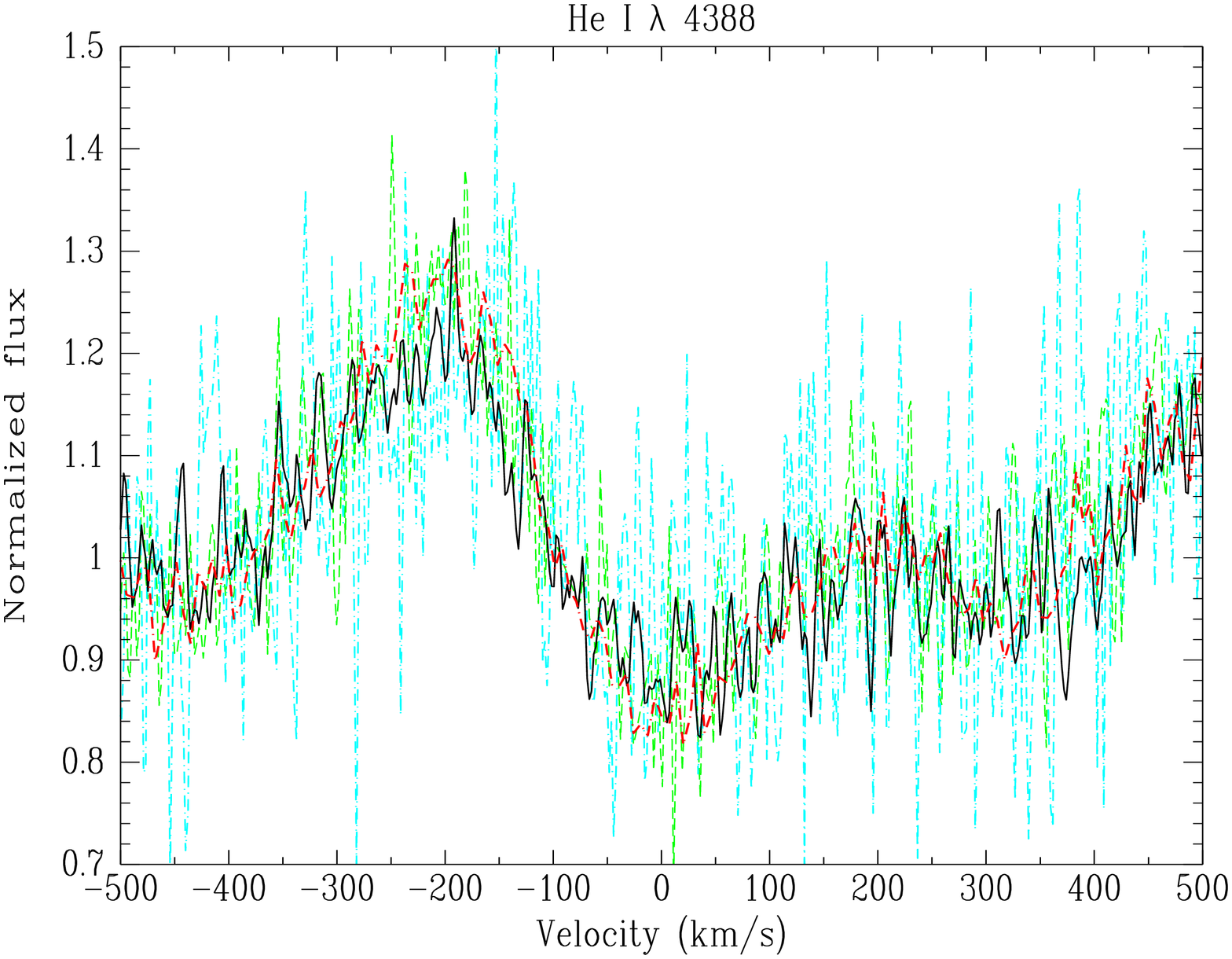}
\includegraphics[angle=270,width=0.5\textwidth]{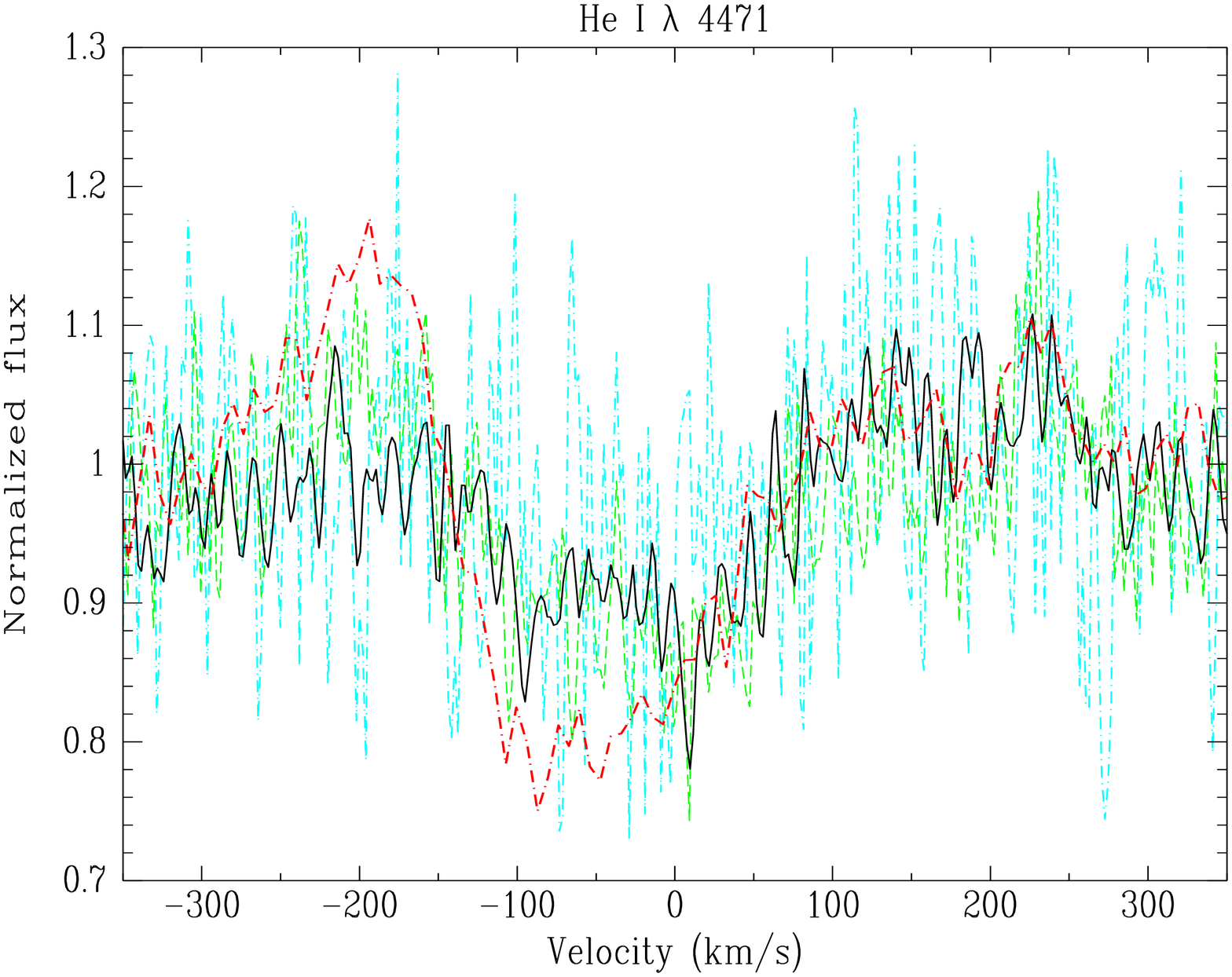}
 \includegraphics[angle=270,width=0.5\textwidth]{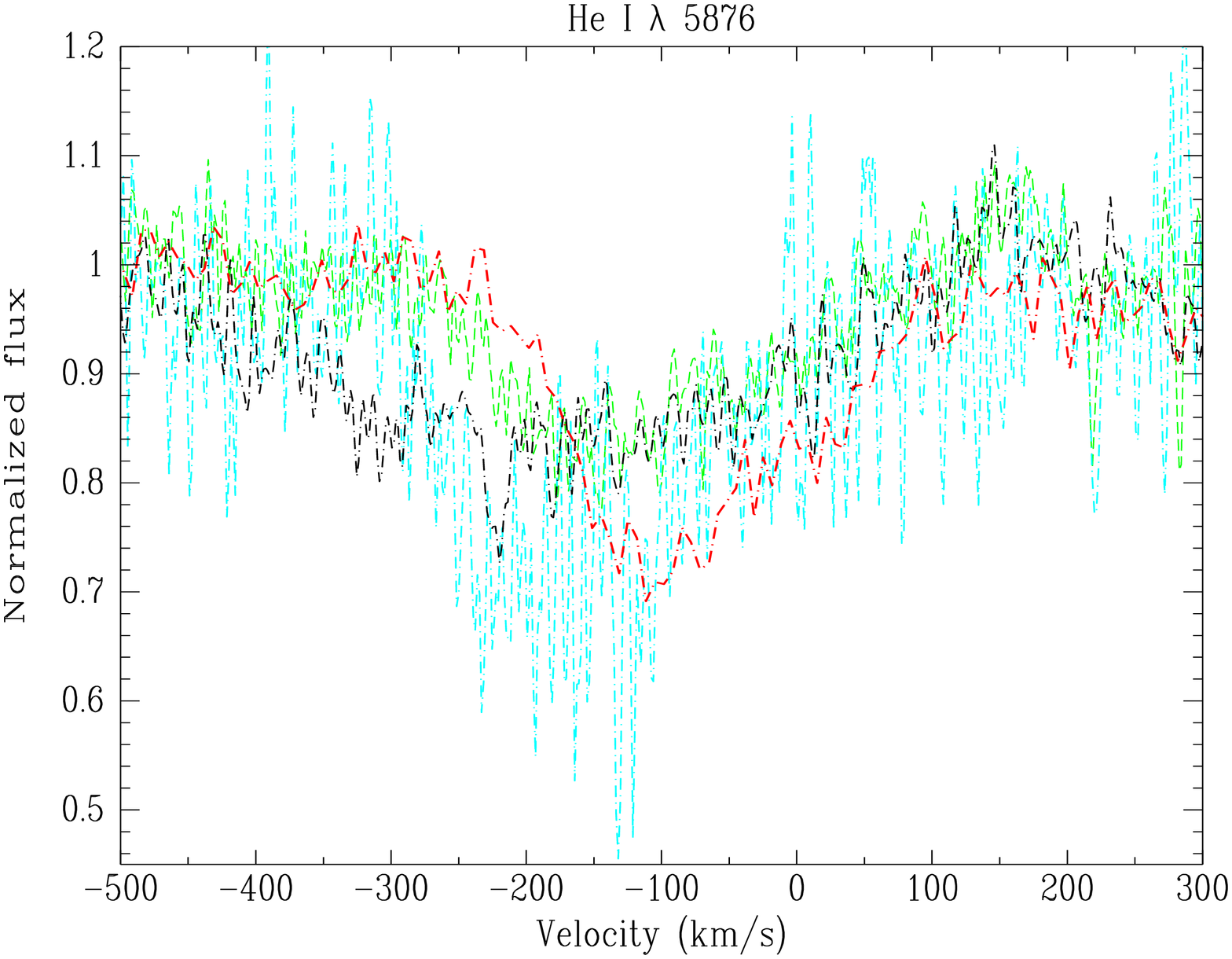}
  \caption{Comparison of the shape of three \ion{He}{i} lines of \object{LHA\,120-S\,35} seen in the du Pont and FEROS spectra on a velocity scale relative to $V_{\rm{sys}}$. Changes in the blue wing of \ion{He}{i} $\lambda\lambda$\,4471 and 5876 lines are clearly observed.}
  \label{Figure-hei-lines}
  \end{centering}
\end{figure}
% -------------------------------------
%______________________________________________________________________________________

An additional pipeline-processed 1-D spectrum was retrieved from the ESO Science Archive Facility, which was obtained with FEROS on 26 November 2015. The exposure time was 2000 s and the final S/N was $\sim$ 11.5 per pixel in the 5500 \AA\, region.

A standard star was observed in each night of 2008 and 2014 for telluric correction, which was performed using standard IRAF tasks. For the FEROS data from 2015 a telluric template from another night was applied.

We complemented these observations with an echelle spectrum ($R \sim$ 12\,600) taken with a REOSC spectrograph attached to the 2.15-m telescope at the Complejo Astron\'omico El Leoncito (CASLEO), Argentina, on 27 November 2012. The adopted instrumental configuration was a 316 l/mm grating, a 350 $\mu$m slit width, and a Tek 1k $\times$ 1k CCD detector, binned in 2$\times$2. The spectrum covers the wavelength range 5800 - 9100 \AA. The exposure time was 2100 s with a resulting S/N ratio per pixel of $\sim$ 6 in the region around 6000 \AA. We followed a standard reduction process using IRAF tasks. As the S/N ratio and the resolution of the CASLEO spectrum are considerably lower than those from the other spectra, we decided to include in this work only the H$\alpha$ region.

\subsection{Near infrared spectra}\label{sec:infrared}

High-quality low resolution spectra ($R \sim$ 1200, S/N  $\sim$ 100) of \object{LHA\,120-S\,35} were taken in the $HK$ bands (12\,500 -- 25\,000 \AA) using the FLAMINGOS-2 spectrograph on the 8.1-m telescope at the Gemini South Observatory (Cerro Pach\'on, Chile) on 21 November 2013. The spectra were taken in the longslit mode using the HK$_G$0806 filter with the HK$_G$5802 grism centered at 1.871 $\mu$m. In order to remove the sky background, an ABBA nodding along the slit was required. A cycle of four ABBA sequences was performed. Immediately after, the late B-type star HIP 24337 (close to the target in position and airmass) was observed for telluric absorption correction. Late B- or early A-type standards are generally selected because the only intrinsic features they exhibit are neutral hydrogen lines. Flats were also acquired for flat fielding. The data were reduced with IRAF/Gemini tasks. We generated a telluric template to correct the science stellar spectrum using the telluric IRAF task. The spectra were wavelength calibrated using the standard telluric star.

%______________________________________________________________________________________
% FIGURE 5-----------------------------
\begin{figure*}
  \begin{centering}
  \includegraphics[angle=270,width=0.5\textwidth]{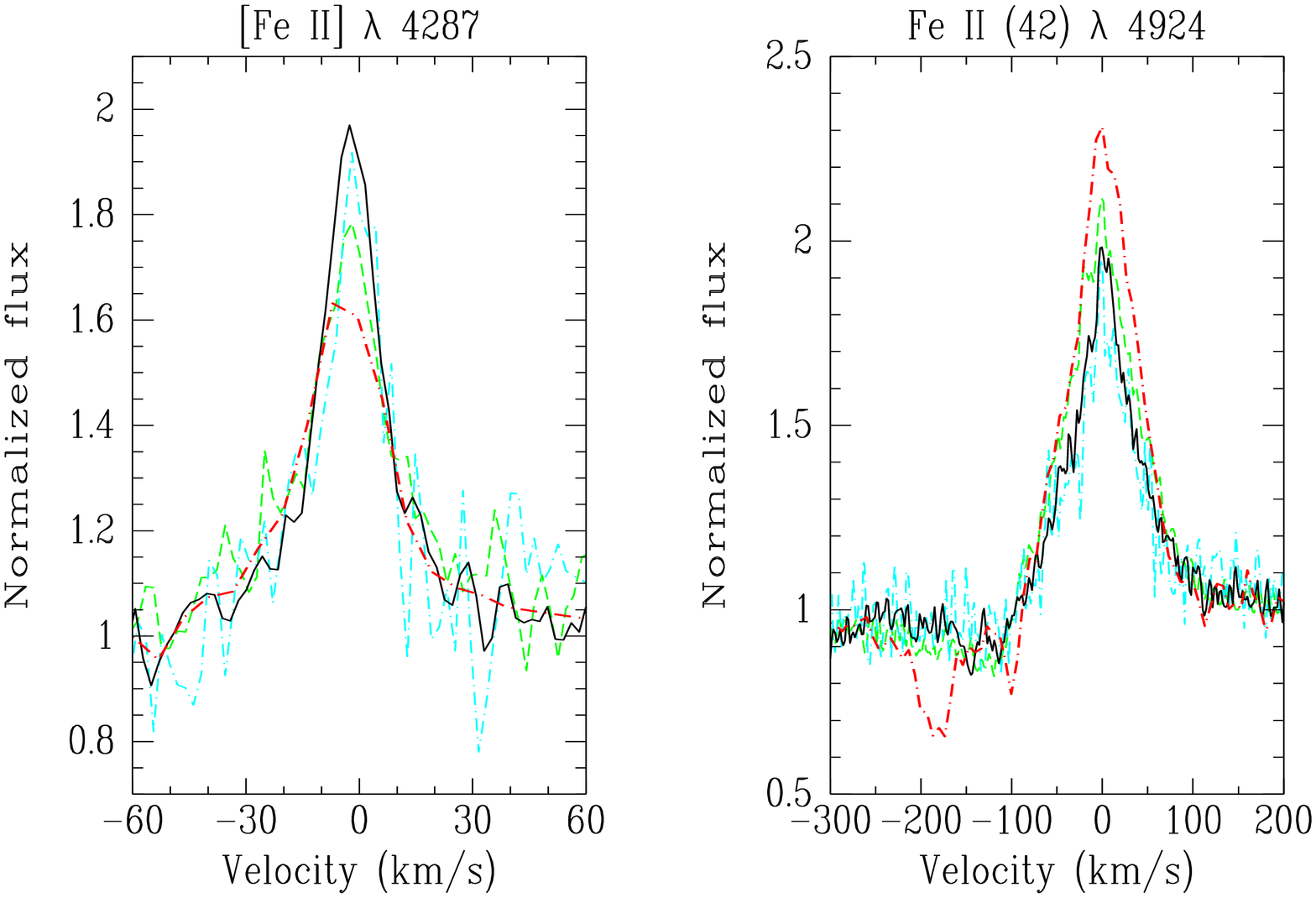},\includegraphics[angle=270,width=0.5\textwidth]{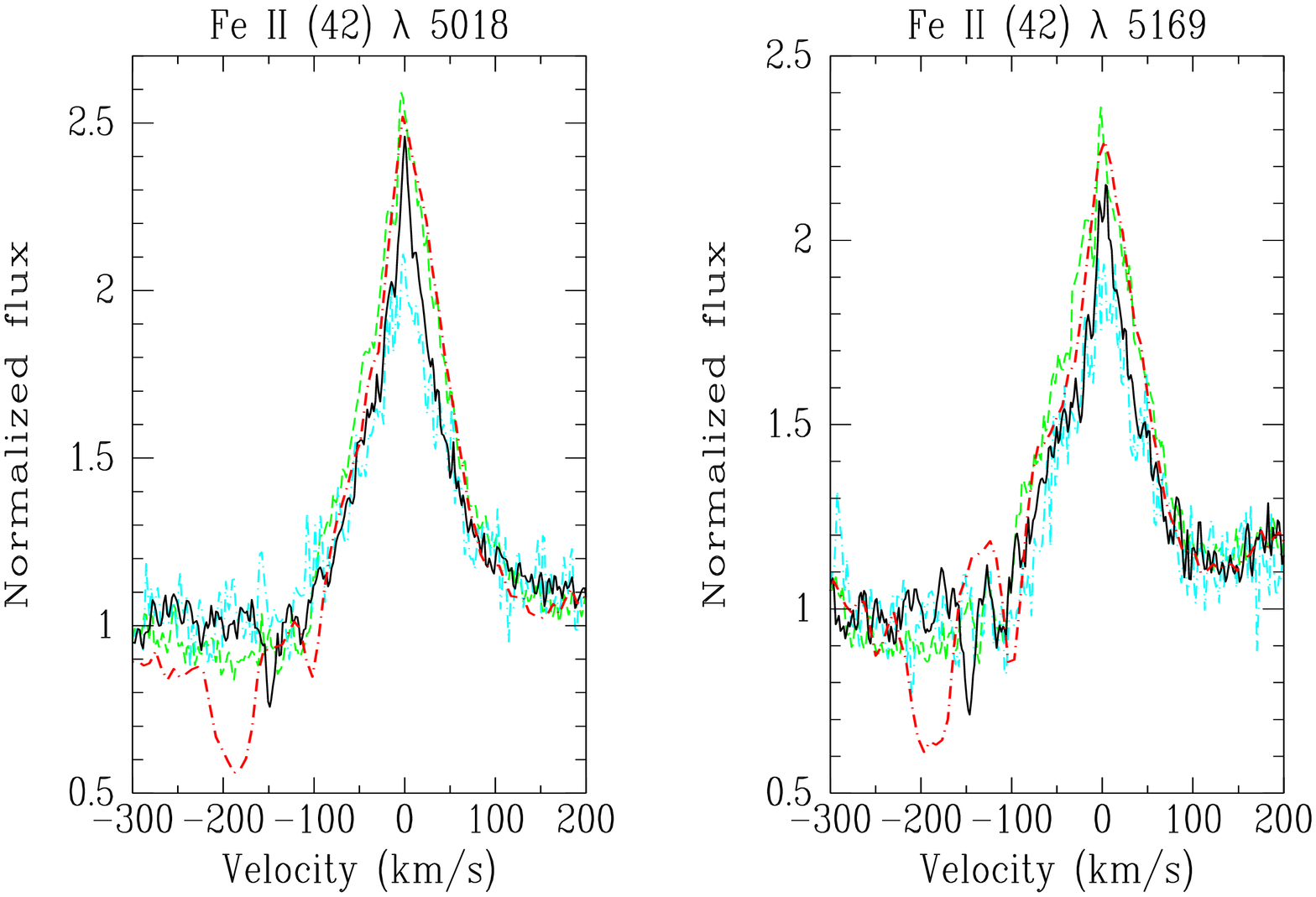}
  \caption{Example of forbidden and permitted \ion{Fe}{ii} lines. The plots show the [\ion{Fe}{ii}] $\lambda$\,4287 line and three permitted lines of the multiplet 42 at $\lambda$\,4924 \AA\, (blended with the \ion{He}{i} $\lambda$\,4922 line), $\lambda$\,5018 \AA\, (blended with the \ion{He}{i} $\lambda$\,5016 line) and $\lambda$\,5169 \AA\, on a velocity scale relative to $V_{\rm{sys}}$. The multiplet 42 transitions display complex P-Cygni profiles, clearly seen in the du Pont and 2015 Nov FEROS spectra.}
  \label{Figure-feii-lines3}
  \end{centering}
\end{figure*}
% -------------------------------------
%______________________________________________________________________________________

An additional observation was secured on 22 November 2012, with OSIRIS (the Ohio State InfraRed Imager/Spectrometer) at the Southern Astrophysical Research (SOAR) 4.1-m telescope (Cerro Pach\'on, Chile). The spectrum was taken in cross-dispersion mode using the f/2.8 camera and a 1024 $\times$ 1024 HAWAII HgCdTe array, covering the spectral range between 12\,500 and 23\,000 \AA\, with a low spectral resolution ($R \sim 1200$). The reduction procedure was done with IRAF tasks, following the same steps than those already mentioned for FLAMINGOS-2 observations.

%________________________________________________________________________________________

% FIGURE 6-----------------------------
\begin{figure}
  \begin{centering}
  \includegraphics[angle=270,width=0.5\textwidth]{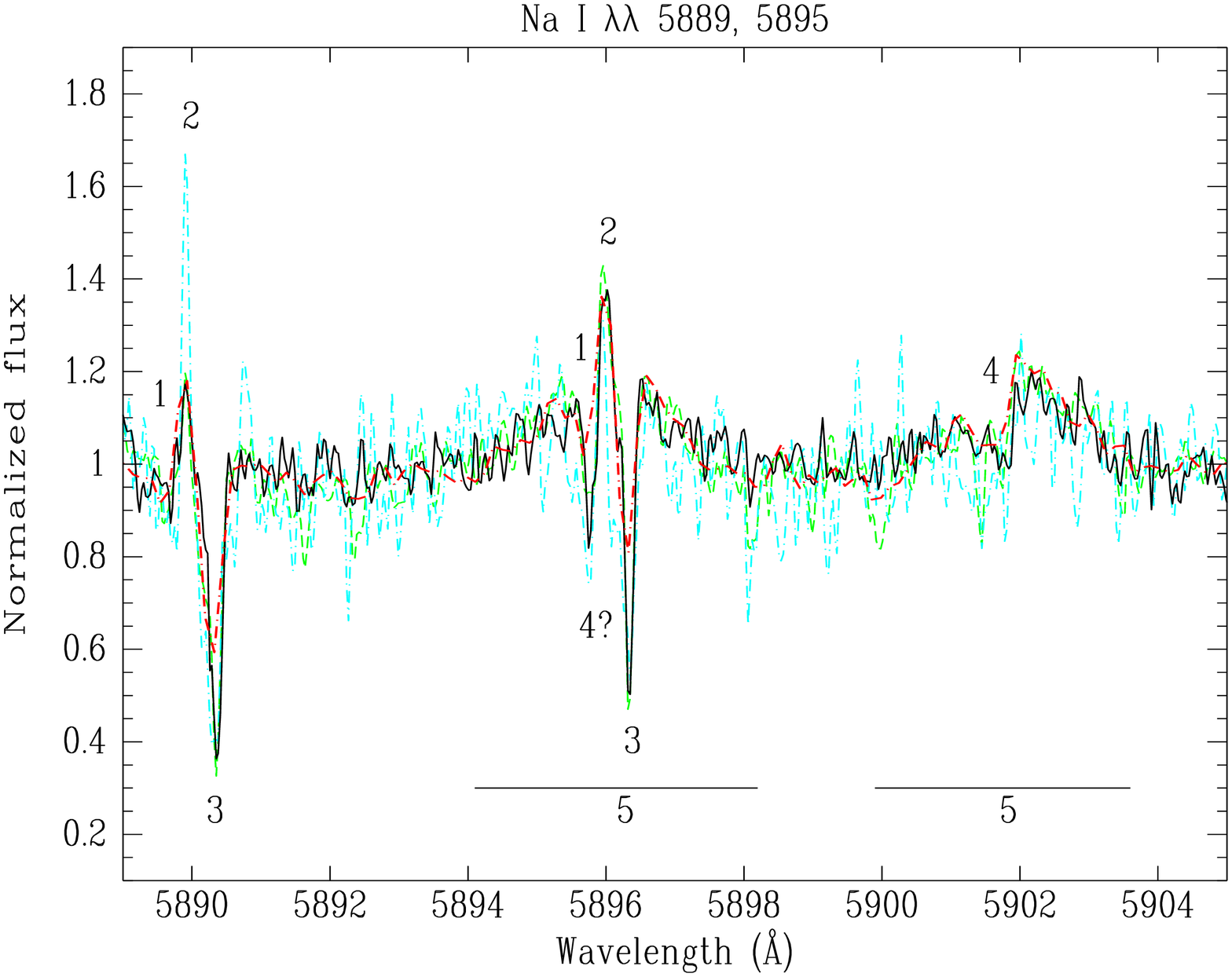}
  \caption{\ion{Na}{i} lines at $\lambda\lambda$\,5889, 5895 \AA\, of \object{LHA\,120-S\,35} seen in the du Pont and FEROS spectra. The doublet has different velocity components related to different formation regions (see text for more details about the labels).} 
  \label{Figure-nai-lines}
 \end{centering}
\end{figure}
% -------------------------------------
%________________________________________________________________________________________

%______________________________________________________________________________________
%______________________________________________________________________________________

\section{Description of the observed main spectral features}\label{sec:results}

\subsection{In the optical}\label{sec:results-visible}

\object{LHA\,120-S\,35} presents a rich optical emission-line spectrum dominated by hydrogen lines as well as permitted and forbidden transitions of singly-ionized elements (such as \ion{Fe}{ii}, \ion{Ti}{ii}, \ion{S}{ii}, \ion{Ca}{ii}) and neutral atoms, like \ion{O}{i}. The only absorption features are the \ion{He}{i} lines. Figures \ref{Figure-3800-5200-lines} and \ref{Figure-7700-9000-lines} show the main spectral features in the ranges of 3800--5200 \AA\, and 7700--9000 \AA\,, respectively.

The blue portion of the spectrum (3800--5200 \AA) resembles the one previously reported by \citet{gum95}. When comparing the same wavelength range (4450--4560 \AA) covered by our observations in 2008 and by \citet[cf. their Fig. 2]{gum95} we find no changes in shape and intensity for the main spectral features.

%________________________________________________________________________________________
% FIGURE 7-----------------------------
\begin{figure*}[!]
  \begin{centering}
  \includegraphics[angle=270,width=0.95\textwidth]{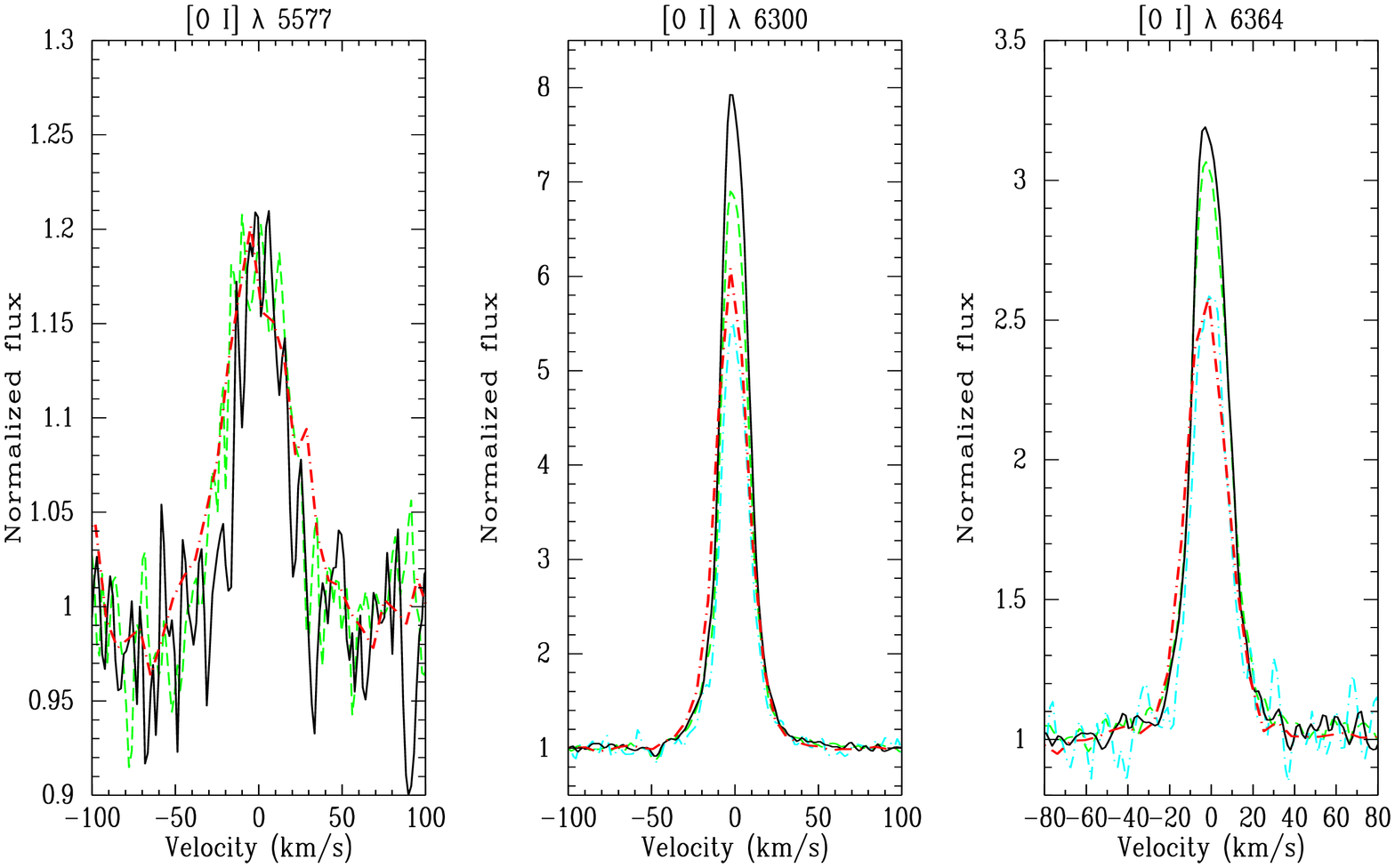}
  \caption{Emission lines of [\ion{O}{i}] at $\lambda\lambda$\,5577, 6300 and 6364 \AA\, of \object{LHA\,120-S\,35} seen in the du Pont and FEROS spectra. We did not include in the analysis the [\ion{O}{i}] $\lambda$\,5577 line corresponding to the FEROS spectrum acquired in Nov 2015 because it is very noisy.}
  \label{Figure-forbidden-oi-lines}
  \end{centering}
\end{figure*}
% -------------------------------------
%________________________________________________________________________________________

{\bf{Hydrogen.}}  Figure \ref{Figure-balmer-lines} shows the high-resolution spectra of the first Balmer lines (H$\alpha$, H$\beta$, H$\gamma$ and H$\delta$) taken with the echelle spectrograph of the du Pont telescope in 2008 and with FEROS in 2014, Nov 2015 and Dec 2015. The spectrum acquired at CASLEO in 2012 is included in the H$\alpha$ plot. The most conspicuous feature of the whole spectrum is the H$\alpha$ line, which shows a double-peaked emission profile with a deep central absorption component at $\sim$ -25 km s$^{-1}$ and a violet-to-red ratio $V/R < 1$, this value being smaller for the Nov 2015 spectrum. The red emission peak is at $\sim$ +21 km s$^{-1}$ and the blue one at $\sim$ -61 km s$^{-1}$, on a velocity scale relative to the systemic velocity $V_{\rm{sys}}$ = +308$\pm$3 km s$^{-1}$, which was derived by \citet{gum95}. A zoom of the H$\alpha$ wings can be seen in Fig. \ref{Figure-halpha-line}. Three narrow absorption components (NACs) are observed on the blue wing of the emission line at around -120 km s$^{-1}$, -226 km s$^{-1}$ and -388 km s$^{-1}$ in the du Pont spectrum and at around -196 km s$^{-1}$, -246 km s$^{-1}$ and -342 km s$^{-1}$ in the FEROS spectrum from 2014 (indicated by arrows in Fig. \ref{Figure-halpha-line}), while in both spectra from 2015 only two dips are noticeable at around -158 km s$^{-1}$ and -277 km s$^{-1}$. In the lower resolution CASLEO spectrum, two NACs are also clearly distinguishable at around -267 km s$^{-1}$ and -359 km s$^{-1}$. These features seem to move across the blue wing of the line.

The profiles of the rest of the first Balmer lines display a complex P-Cygni structure with its blue edge reaching a value of up to -438 km s$^{-1}$. Similar to the H$\alpha$ line, the emission component presents two peaks at a mean wavelength of $\sim$ +12 km s$^{-1}$ and $\sim$ -40 km s$^{-1}$. The blue-shifted central absorption moves to higher velocities as we consider lower members of the series. In addition, NACs are detectable, especially in the spectrum from 2014, superimposed on the bottom of the profile's absorption trough of the H$\beta$ line at $\sim$ -151 km s$^{-1}$ and $\sim$ -322 km s$^{-1}$ and of the H$\gamma$ and H$\delta$ lines at $\sim$ -143 km s$^{-1}$, $\sim$ -223 km s$^{-1}$ and $\sim$ -301 km s$^{-1}$. The high members of the Balmer series exhibit typical P-Cygni features which are observed up to H$_{22}$. Regarding the medium-resolution spectrum published by \citet{gum95}, the overall shape and intensity of the Balmer lines are the same as in the du Pont spectra.

Hydrogen Paschen lines are observed in pure emission up to n = 31 with a velocity relative to $V_{\rm{sys}}$ close to zero. The full width at half maximum (FWHM) of the lines ranges from 75 km s$^{-1}$ to 165 km s$^{-1}$, the spectrum from Nov 2015 presenting the widest lines. The line intensities in the spectrum from 2014 are higher than the ones from 2008 and 2015 (see Fig. \ref{Figure-caii-lines}), particularly the ones from Nov 2015 are the lowest. However, the intensities of the Paschen lines turn to be similar at the end of the series (see Fig. \ref{Figure-paschen-lines}), indicating that these lines form in denser and, therefore, inner regions of the envelope.

{\bf{Helium.}} The \ion{He}{i} lines are in absorption. The \ion{He}{i} $\lambda$\,4388 singlet shows a complex structure (blended with the \ion{Ti}{ii} $\lambda$\,4387 transition) and remains unchanged throughout all epochs, even in the poorest quality spectrum from Nov 2015 (where no global changes are observed whatsoever). Conversely, noticeable variations are observed in the line depth and the shape of the blue wing of \ion{He}{i} $\lambda$\,4471 and $\lambda$\,5876 triplets (see Fig. \ref{Figure-hei-lines}). The \ion{He}{i} $\lambda$\,4471 line observed in the spectrum of the year 2008 resembles the one reported by \citet{gum95}, who identified the emission feature seen in the blue wing of the \ion{He}{i} line as the \ion{Ti}{ii} $\lambda$\,4468 transition. The change in the blue wing of the \ion{He}{i} $\lambda$\,5876 line profile is clearly detected in all spectra, even between the 2015 FEROS spectra obtained approximately one month apart. The \ion{He}{i} $\lambda$\,4922 and $\lambda$\,5016 singlets are blended with the complex P-Cygni profiles of \ion{Fe}{ii} (42) multiplet (see Fig. \ref{Figure-feii-lines3}), whose complex structure is blurred in the 2015 FEROS spectra. None of our spectra displays the \ion{He}{ii} $\lambda$\,4686 line. 

{\bf{Iron.}} The spectrum is rich in both \ion{Fe}{ii} and [\ion{Fe}{ii}] emission lines. The FWHMs of \ion{Fe}{ii} and [\ion{Fe}{ii}] lines are about 85--105 km s$^{-1}$  and 17--25 km s$^{-1}$, respectively. In Fig. \ref{Figure-feii-lines3}, the [\ion{Fe}{ii}] $\lambda$\,4287 line and three \ion{Fe}{ii} transitions of multiplet 42 are shown. The last ones show complex P-Cygni features (clearly seen in the du Pont spectra and Nov 2015 FEROS spectra) with remarkable variations in the intensity of the absorption component. 

{\bf{Sodium.}} Several components are identified in the D1 and D2 lines (see Fig. \ref{Figure-nai-lines}). The absorption components 1 and 3 could be originated in the galactic interstellar medium (ISM). Their apparent complex feature could be the result of their blending with the night sky lines (components 2). Two broad components in emission (components 5) with a radial velocity similar to the systemic velocity, $\sim$308 km s$^{-1}$, are also seen, probably arising from the circumstellar environment; their profiles do not show temporal variations. Finally, an absorption component (component 4) at $\sim$277 km s$^{-1}$ is clearly detected superimposed to the second broad emission feature (D2 line). Since its radial velocity is close to the velocity of the LMC, this component 4 could be related to the LMC interstellar medium. The corresponding component 4 of the D1 line could be blended with the complex galactic ISM feature at $\lambda$\,5895 \AA\, (component 1).

%________________________________________________________________________________________
% FIGURE 8-----------------------------
\begin{figure}[!th]
  \begin{centering}  
  \includegraphics[angle=270,width=0.5\textwidth]{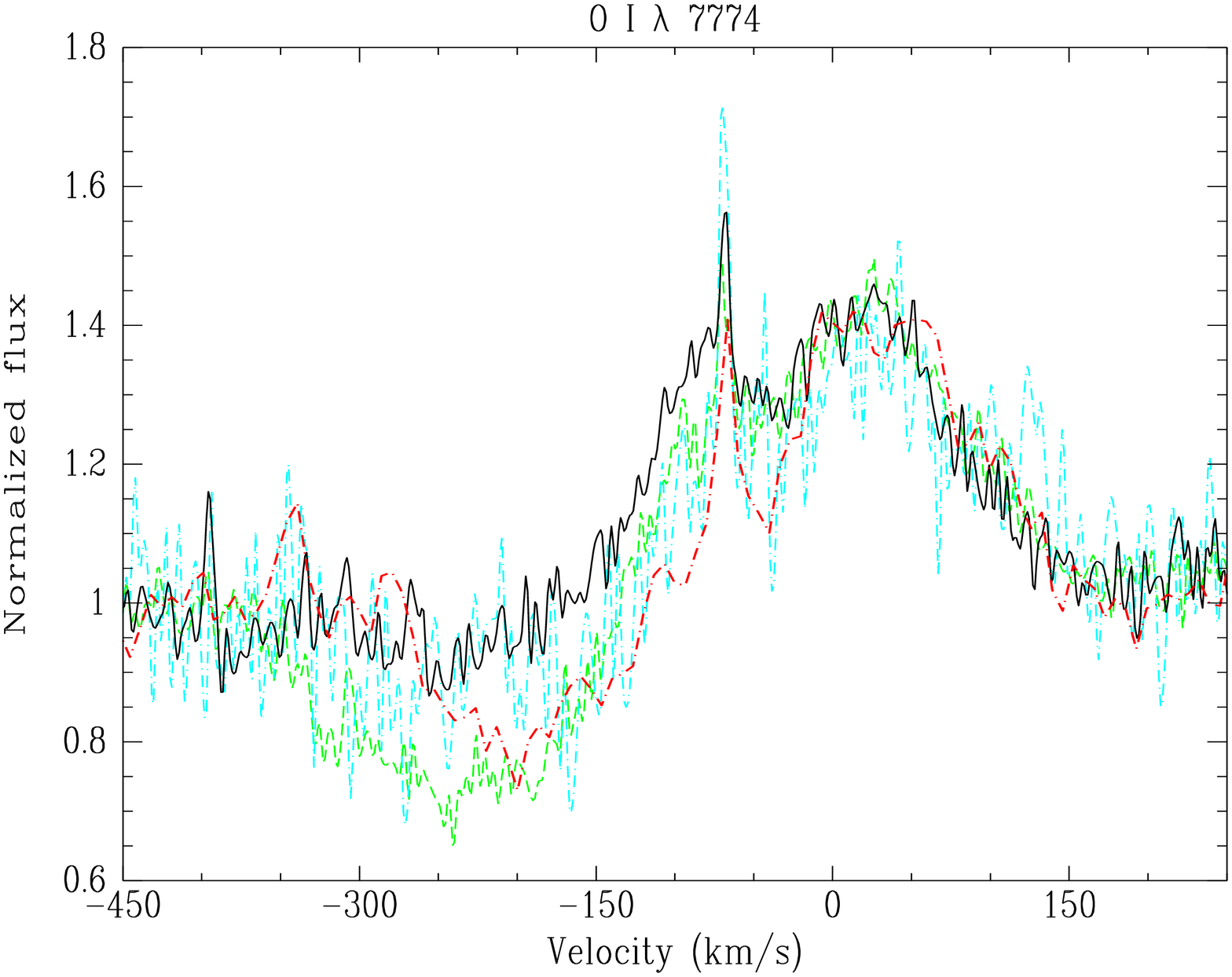}
  \includegraphics[angle=270,width=0.5\textwidth]{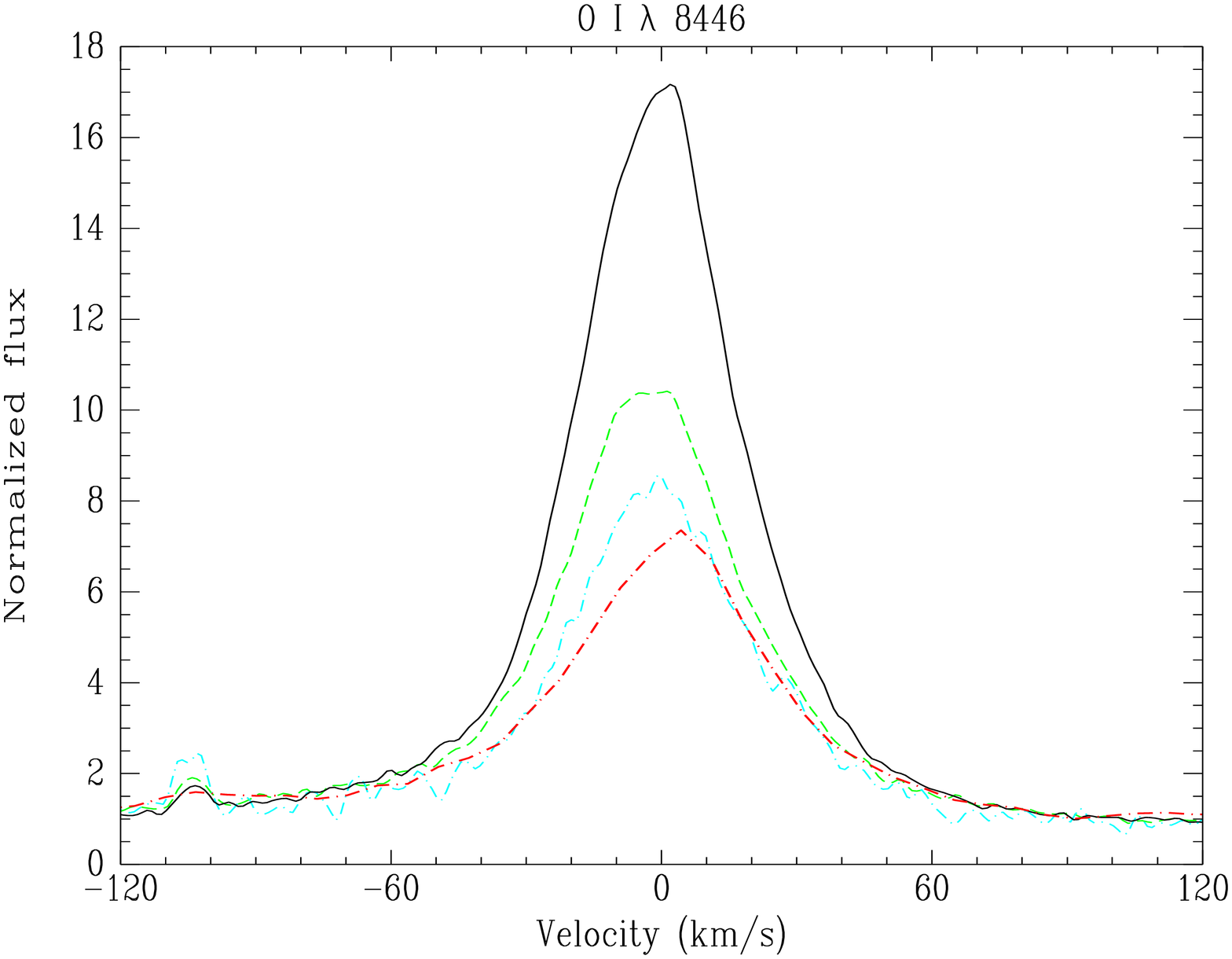}
  \caption{Top panel: Line profile of the \ion{O}{i} $\lambda$\,7772-75 triplet on a velocity scale relative to $V_{\rm{sys}}$. Bottom panel: Comparison of the strength and shape of the permitted transition of \ion{O}{i} $\lambda$\,8446.}
  \label{Figure-oi-lines}
  \end{centering}
\end{figure}
% -------------------------------------
% FIGURE 9-----------------------------
\begin{figure}[!ht]
  \begin{centering}  
  \includegraphics[angle=270,width=0.5\textwidth]{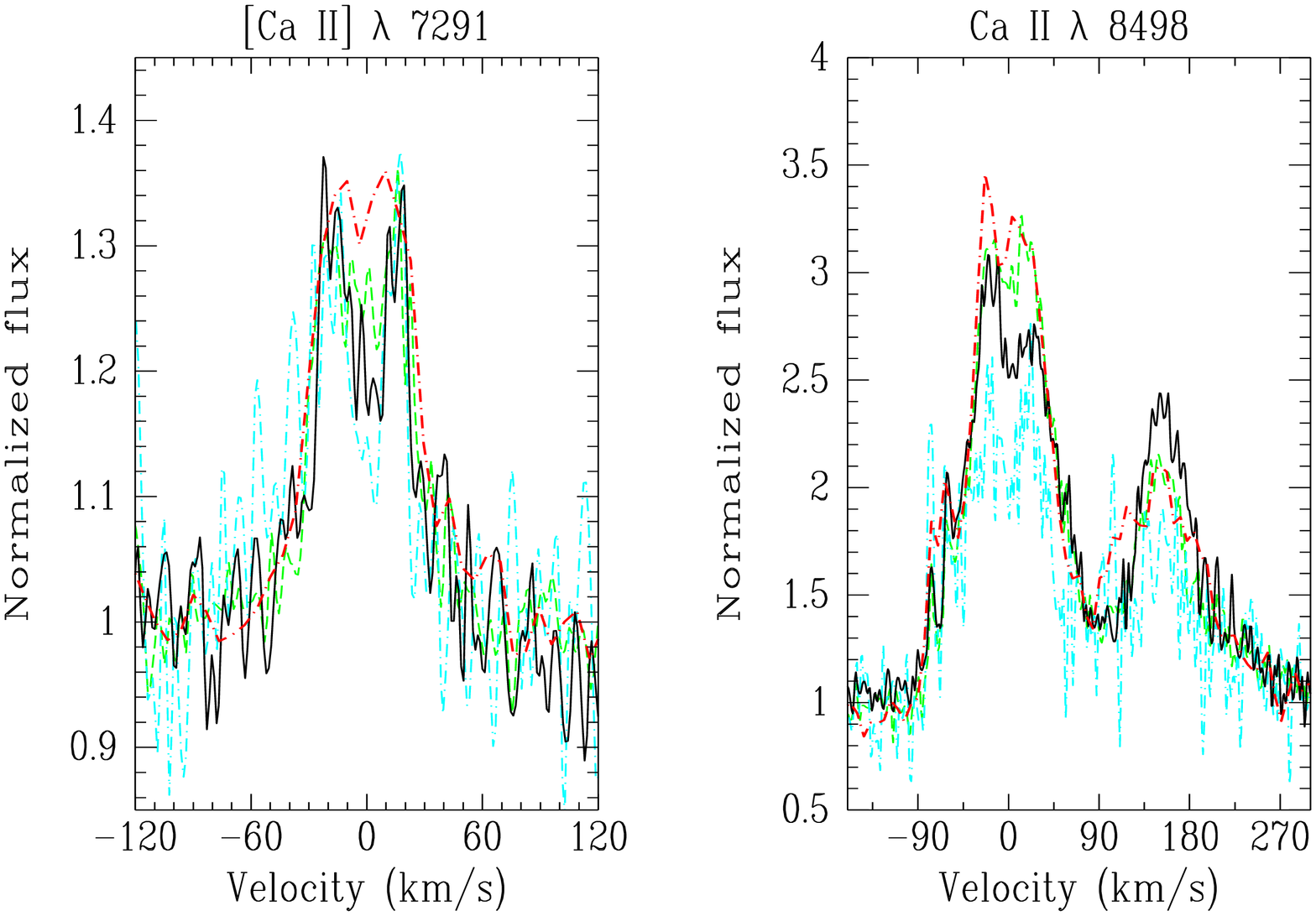}
  \includegraphics[angle=270,width=0.5\textwidth]{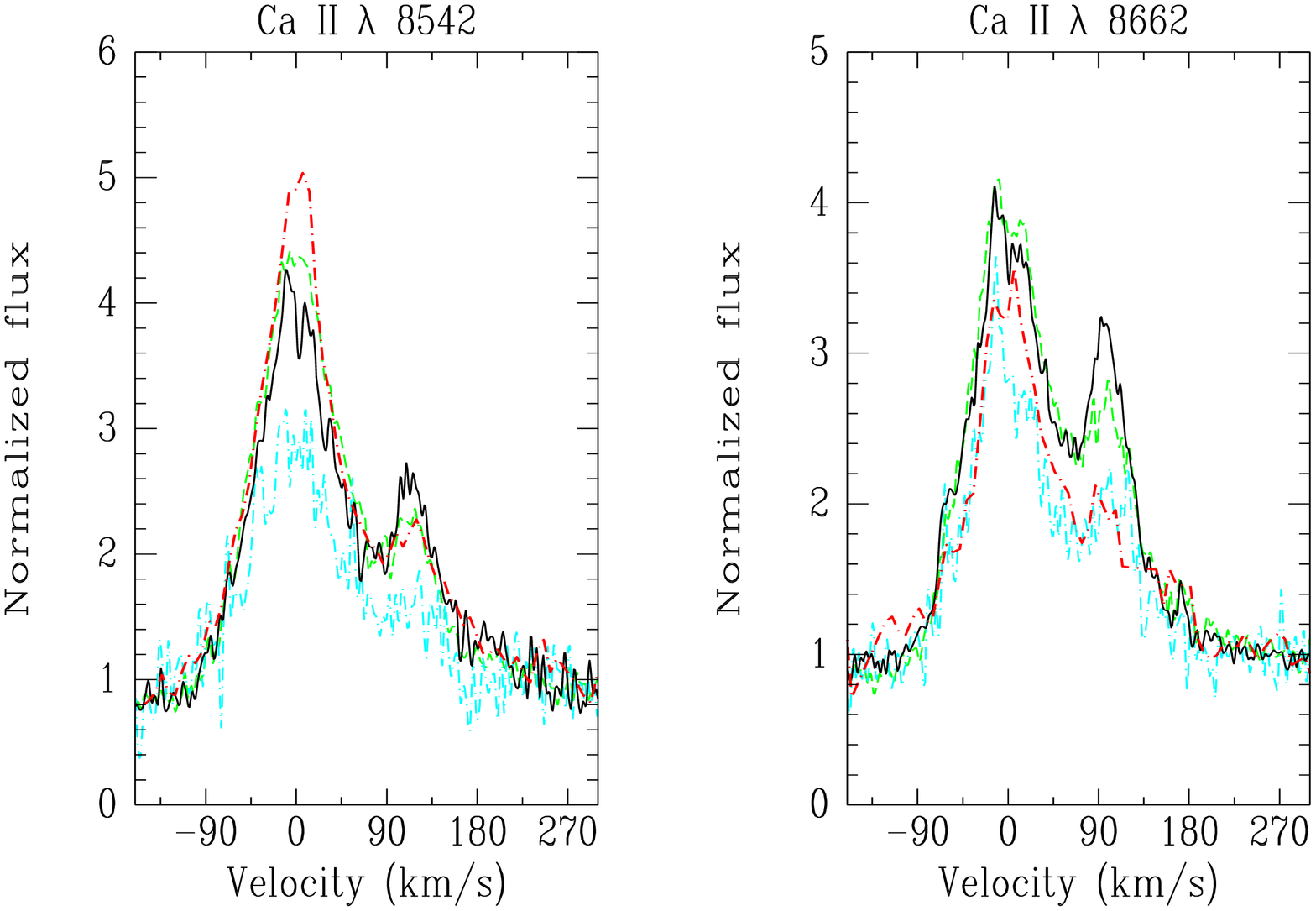}
  \caption{Comparison of the strength and shape of the [\ion{Ca}{ii}] $\lambda$\,7291 and the \ion{Ca}{ii} $\lambda\lambda$\,8498, 8542, 8662 emission lines seen in the du Pont data and in FEROS spectra from 2014 and 2015. The \ion{Ca}{ii} triplet lines are blended with the Paschen lines Pa16, Pa15 and Pa13, respectively.}
  \label{Figure-caii-lines}
  \end{centering}
\end{figure}
% -------------------------------------
% FIGURE 10-----------------------------
\begin{figure}[!ht]
  \begin{centering}  
  \includegraphics[angle=270,width=0.4\textwidth]{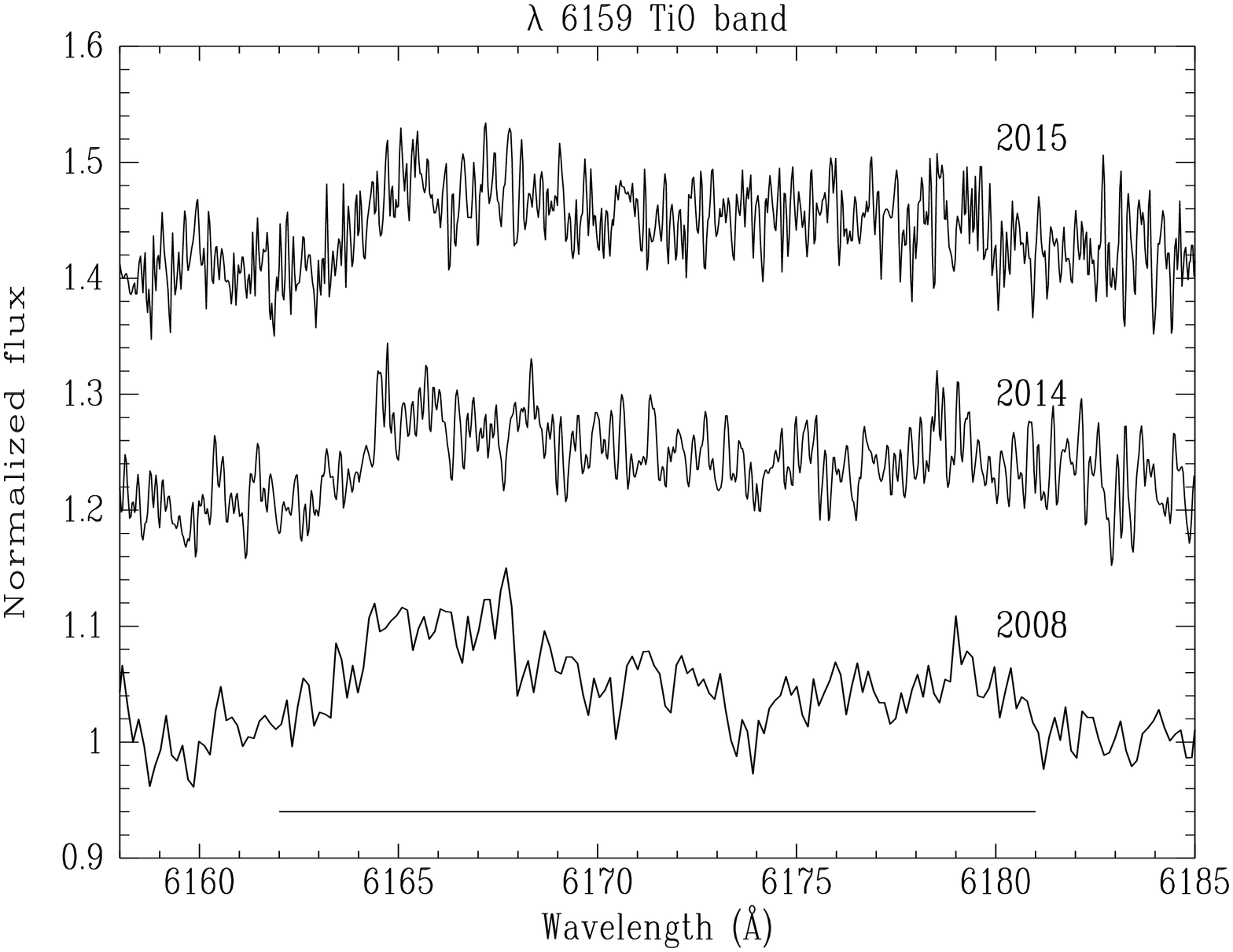}
  \caption{Detection of a weak TiO band emission in the high-resolution optical spectra of \object{LHA\,120-S\,35}. The TiO feature from the Nov 2015 data is not included since it is very noisy.}
  \label{Figure-tio-bands}
  \end{centering}
\end{figure}
% -------------------------------------
%________________________________________________________________________________________
 
{\bf{Oxygen.}} The spectra show prominent single-peaked emission lines of [\ion{O}{i}] $\lambda\lambda$\,6300, 6364 \AA\, with FWHMs of $\sim$ 24 km s$^{-1}$ in the du Pont data and $\sim$ 19 km s$^{-1}$ in the FEROS observations; these lines in the spectrum taken in 2014 are the most intense (see Fig. \ref{Figure-forbidden-oi-lines}). The [\ion{O}{i}] $\lambda$\,5577 line is also visible and it seems not to show sensitive variations in strength (we excluded from the analysis the one corresponding to the Nov 2015 spectrum because it is very noisy). However, its FWHM ranges from $\sim$ 28 km s$^{-1}$ in the FEROS spectrum from 2014 to $\sim$44 km s$^{-1}$ in the rest of the spectra. The well-known forbidden lines of \ion{O}{iii} at $\lambda\lambda$\,4959 and 5007 \AA\, are absent. The strongest line of [\ion{O}{ii}] at $\lambda$\,7319 \AA\, is also not present.

%________________________________________________________________________________________
% FIGURE 11-----------------------------
\begin{figure*}[!t]
  \begin{centering}
  \includegraphics[angle=270,width=1.\textwidth]{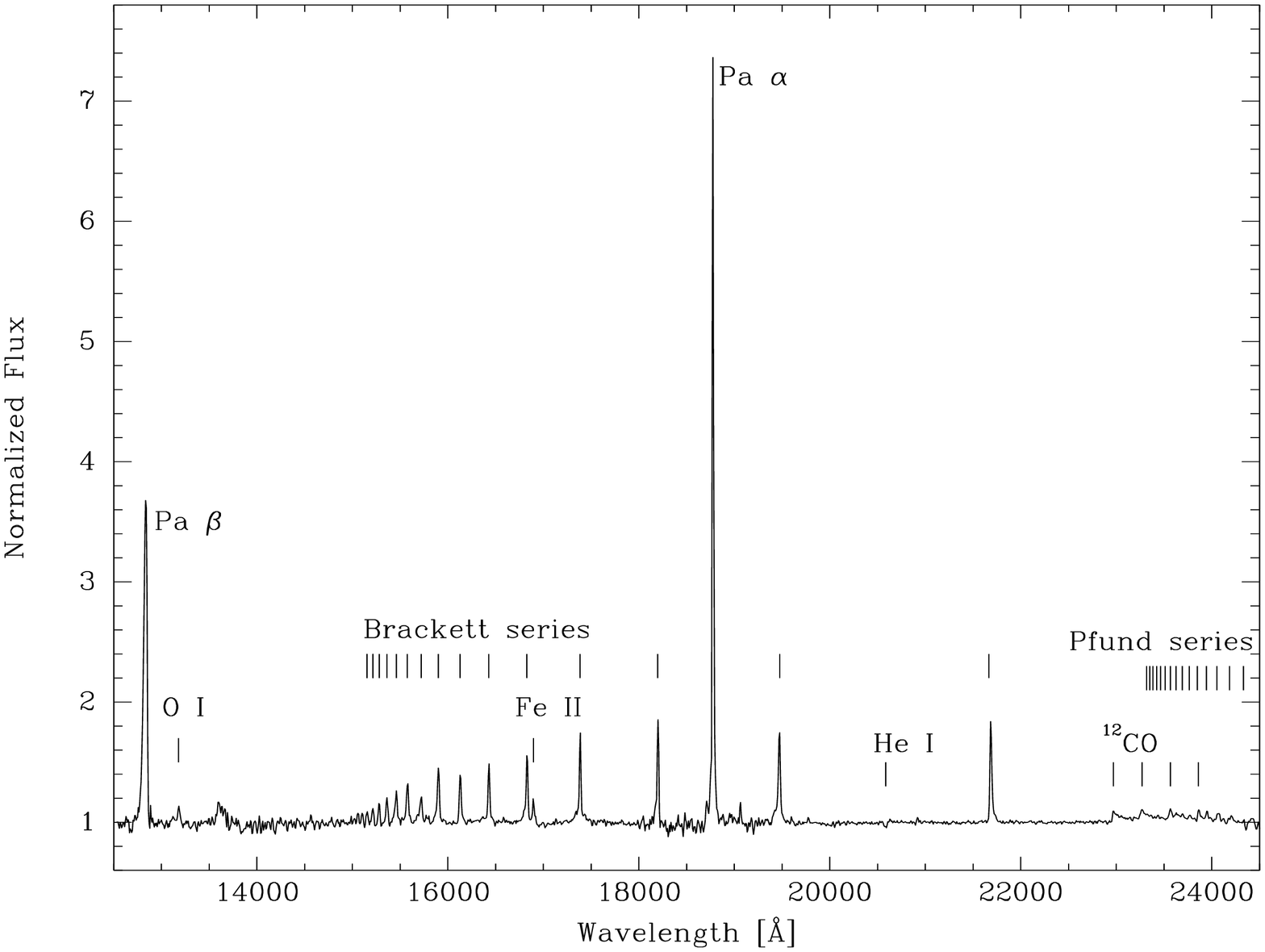}
  \caption{Flux-normalized infrared spectrum of \object{LHA\,120-S\,35} observed with the FLAMINGOS-2 spectrograph. The plot shows the Pa$\beta$ and Pa$\alpha$ prominent emission lines. The hydrogen Brackett series, the band heads of $^{12}$CO and some other individual lines are indicated.}
  \label{Figure-ir-spectrum-flamingos2} 
  \end{centering}
\end{figure*}
% --------------------------------------
%________________________________________________________________________________________

In the red spectral region, the permitted \ion{O}{i} $\lambda$\,8446 triplet is the most remarkable emission feature, which could be attributed to a Ly $\beta$ fluorescence mechanism \citep{jas93,mat12}. It is slightly blended with the Pa18 line ($\lambda$ 8438 \AA\,). Between the years 2008 and 2014, the line has increased its intensity by more than a factor of two, decreasing in the following year to about 40\,\% of the maximum value and increasing again by more than 10\,\% of this value in approximately one month. On the other hand, the \ion{O}{i} triplet at $\lambda$\,7772-75 \AA\, shows a P-Cygni line profile with a weak emission. Figure \ref{Figure-oi-lines} shows the variation in the P-Cygni absorption component of the triplet, which in the FEROS spectra of 2014 and Nov 2015 appears partially refilled by emission. 

{\bf{Calcium.}} The forbidden transitions of \ion{Ca}{ii} at $\lambda\lambda$\,7291, 7324 \AA\, are clearly detected in emission. The [\ion{Ca}{ii}] $\lambda$\,7291 line is variable and presents a double-peaked profile with a peak separation of $\sim$ 21 km s$^{-1}$ from du Pont spectrum and an average of $\sim$ 33 km s$^{-1}$ from FEROS spectra, with a notable width ranging from $\sim$ 110 to 50 km s$^{-1}$ in average, respectively (top left panel of Fig. \ref{Figure-caii-lines}). Since [\ion{Ca}{ii}] doublet arises in a region of strong telluric pollution, it is difficult to obtain reliable measurements, especially if the lines are weak and the tellurics cannot properly be removed; such is the case of the [\ion{Ca}{ii}] $\lambda$\,7324 line (not shown here).

The near-infrared triplet of \ion{Ca}{ii} $\lambda\lambda$\,8498, 8542, 8662 presents strong double-peaked emission features blended with the Paschen lines at $\lambda$\,8502 \AA\,, $\lambda$\,8545 \AA\, and $\lambda$\,8665 \AA\,, respectively (see Fig. \ref{Figure-caii-lines}). While hydrogen lines exhibit a noticeable intensity variation in 2014 with respect to the spectrum from 2008, the \ion{Ca}{ii} triplet shows a small intensity change, except for the weak calcium profiles registered in Nov 2015.

{\bf{Nitrogen and sulphur}}. The [\ion{N}{ii}] $\lambda$\,5754 and [\ion{N}{ii}] $\lambda$\,6584 lines are also present in emission, but the [\ion{N}{ii}] $\lambda$\,6548 line is only seen in the spectra acquired in 2014 and 2015 (see Fig. \ref{Figure-halpha-line}). The \ion{N}{iii} triplet at $\lambda\lambda$\,4634, 4640, 4642 \AA\, is absent in all our spectra.
The transitions of [\ion{S}{ii}] at $\lambda\lambda$\,4069, 4076 \AA\, are present in emission in all spectra as well as the transitions at $\lambda\lambda$\,6717, 6731\,\AA\,, although the last ones are very weak.

{\bf{Titanium oxide}}. From $\sim$ 6162 \AA\, to 6181 \AA\, a broad and weak emission feature can be detected in all epochs (see Fig. \ref{Figure-tio-bands}) with similar strength. In previous works a similar structure, clearly observed in four B[e]SGs belonging to the Magellanic Clouds, was associated to a TiO band emission \citep{zic89, tor12, kra16}.

\subsection{In the near-infrared}

$HK$-band spectra of \object{LHA\,120-S\,35} were taken with the OSIRIS and FLAMINGOS-2 spectrographs on November 2012 and 2013, respectively. The good-quality FLAMINGOS-2 spectrum displays the hydrogen Pa$\alpha$ and Pa$\beta$ lines in a prominent emission (see Fig. \ref{Figure-ir-spectrum-flamingos2}). The Brackett series lines are also visible in emission, up to the $n$ = 22 transition.

Moreover, the FLAMINGOS-2 spectrum shows the {\ion{O}{i}} $\lambda$\,1.316 $\mu$m and the {\ion{Fe}{ii}} $\lambda$\,1.687 $\mu$m lines. A weak absorption feature is also observed near 2.058 $\mu$m that corresponds to {\ion{He}{i}}.  Around the Pa$\alpha$ line, the spectrum appears noisy, which could be related to a residual of the telluric correction process since at 1.9 $\mu$m there is an intense telluric water band. In the same way, the broad emission feature at 1.36 $\mu$m could be a result of the difficulty in removing the intense telluric band at 1.4 $\mu$m completely.

%________________________________________________________________________________________
% FIGURE 12-----------------------------   
\begin{figure}[!b]
  \begin{centering}
  \includegraphics[angle=-90,width=0.5\textwidth]{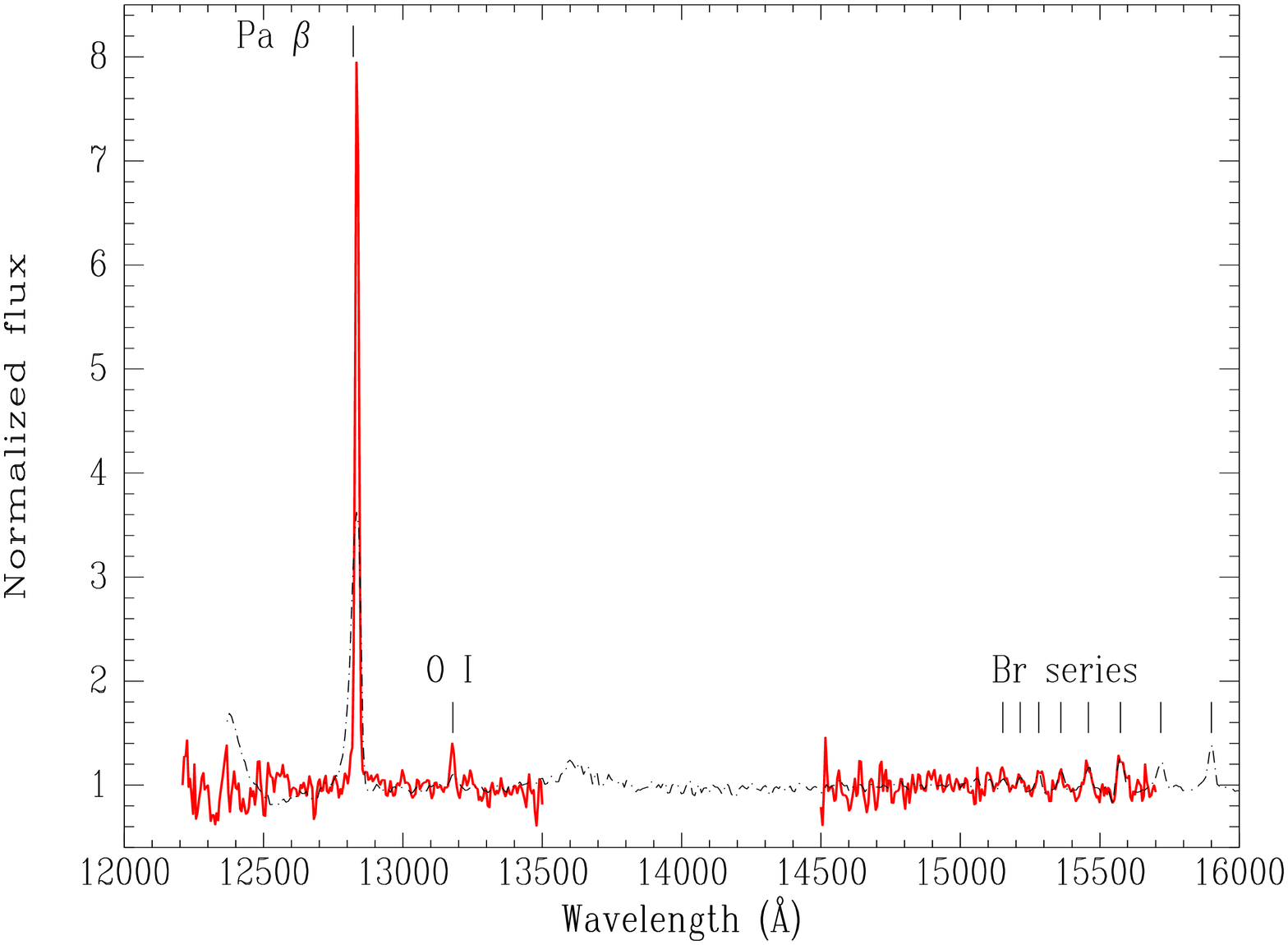}
  \includegraphics[angle=-90,width=0.5\textwidth]{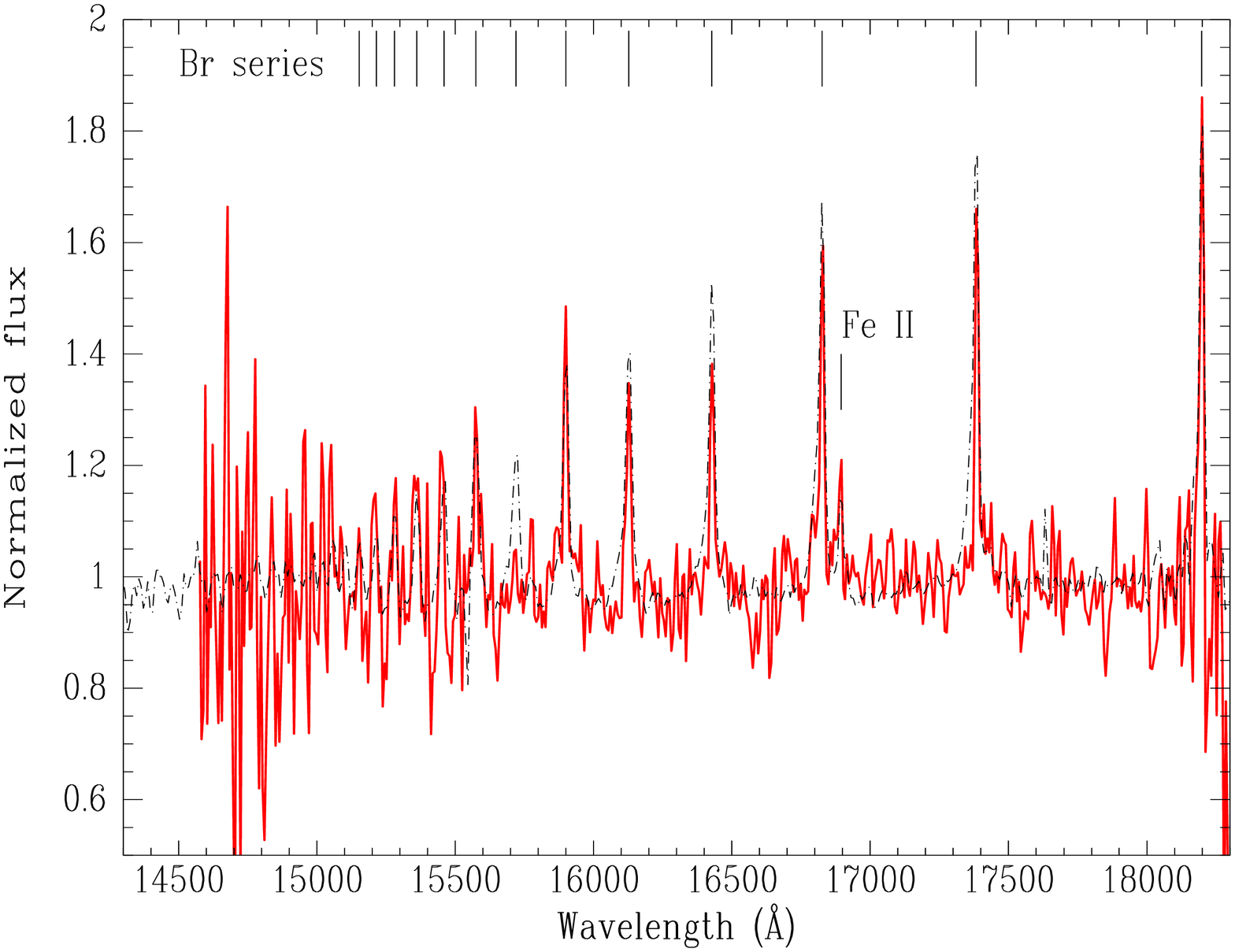}
  \includegraphics[angle=-90,width=0.5\textwidth]{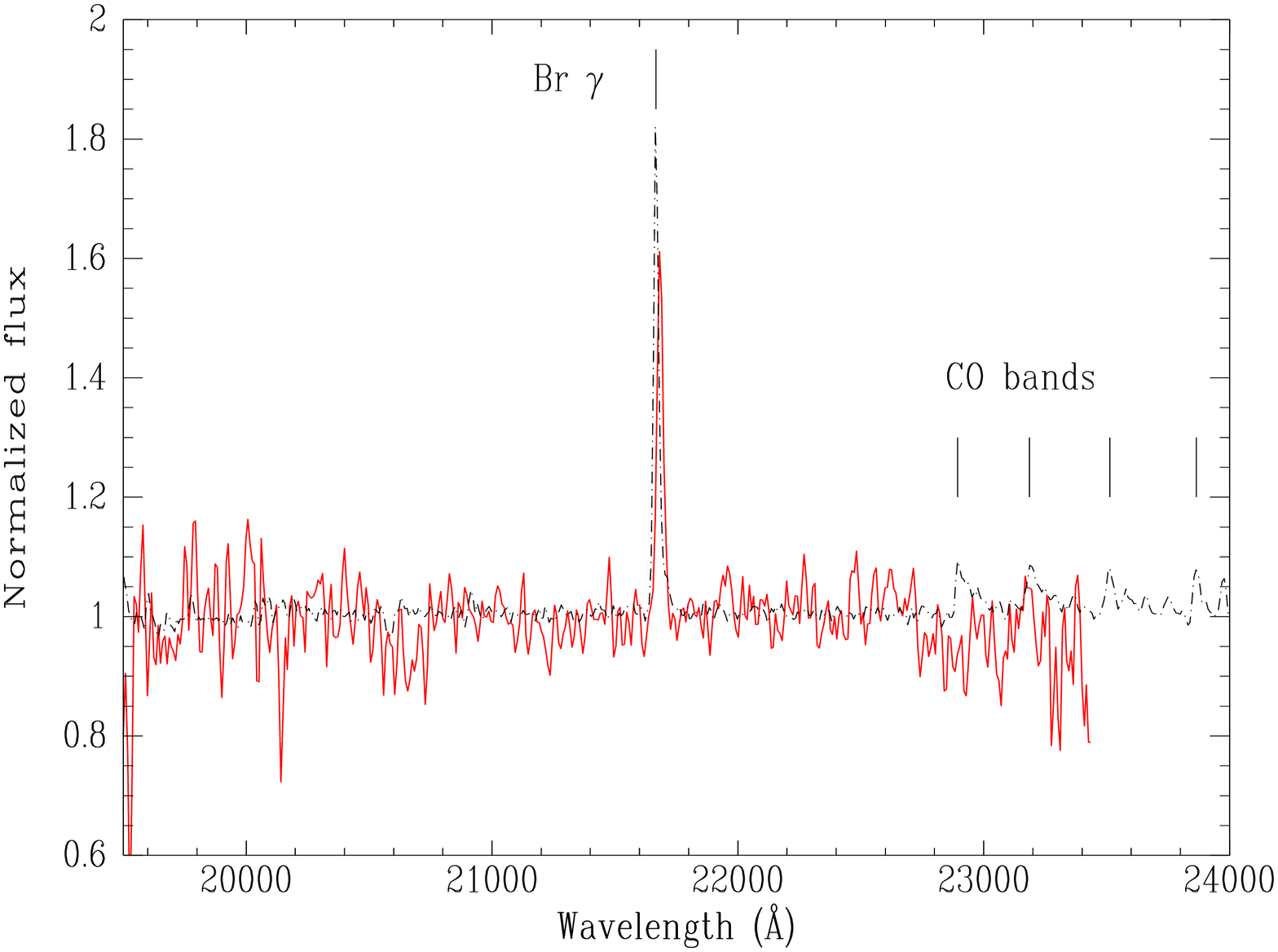}
  \caption{Comparison of the flux-normalized infrared spectra of \object{LHA\,120-S\,35} taken with the OSIRIS (in red solid line) and FLAMINGOS-2 (in dash-dotted black line) spectrographs. Despite of the poor S/N ratio of the OSIRIS data, it can be seen that while the Pa$\beta$ line is more than twice stronger than in the FLAMINGOS-2 spectrum, the Brackett series emission lines are weaker.}
 \label{Figure-ir-spectrum-osiris-flamingos2}
 \end{centering}
\end{figure}
% -------------------------------------
%________________________________________________________________________________________

Additionally, weak CO first overtone band emission is detected where the following $^{12}$CO band heads can be identified: 2-0 (2.294 $\mu$m), 3-1 (2.323 $\mu$m), 4-2 (2.352 $\mu$m), 5-3 (2.383 $\mu$m). Weak emission of the hydrogen Pfund series is observed in the CO band region as well.

Strong variability in the strength of the hydrogen lines and CO bands is detected when comparing the OSIRIS to the FLAMINGOS-2 spectra, taken almost one year later. Figure \ref{Figure-ir-spectrum-osiris-flamingos2} shows that in spite of the poor quality of the OSIRIS spectrum the intensity of the Brackett lines is weaker than that observed in the spectrum of FLAMINGOS-2 but conversely the Pa$\beta$ line appears more than twice stronger. Regarding the CO bands in the OSIRIS spectrum, there does not seem to be any clear signs of emission. We cannot assert if the emission is either extremely weak or absent due to both the low resolution of the spectrum  and the poor telluric correction beyond 2.2 $\mu$m (see Fig. \ref{Figure-ir-spectrum-osiris-flamingos2}).

%______________________________________________________________________________________
%______________________________________________________________________________________

\section{Circumstellar gas variability}\label{sec:kinematics}

We collected optical and near-IR spectra of high, moderate and low spectral resolution of \object{LHA\,120-S\,35}. All this information allowed us to trace the kinematical properties of atomic and molecular gaseous emitting regions and discuss their variability. We assumed that the ejected mass from the star accumulates in a disc-like circumstellar structure and analysed its properties considering a Keplerian rotating disc.

%______________________________________________________________________________________
% FIGURE 13-----------------------------
\begin{figure*}[]
  \begin{centering}
  \includegraphics[angle=0,width=0.33\textwidth]{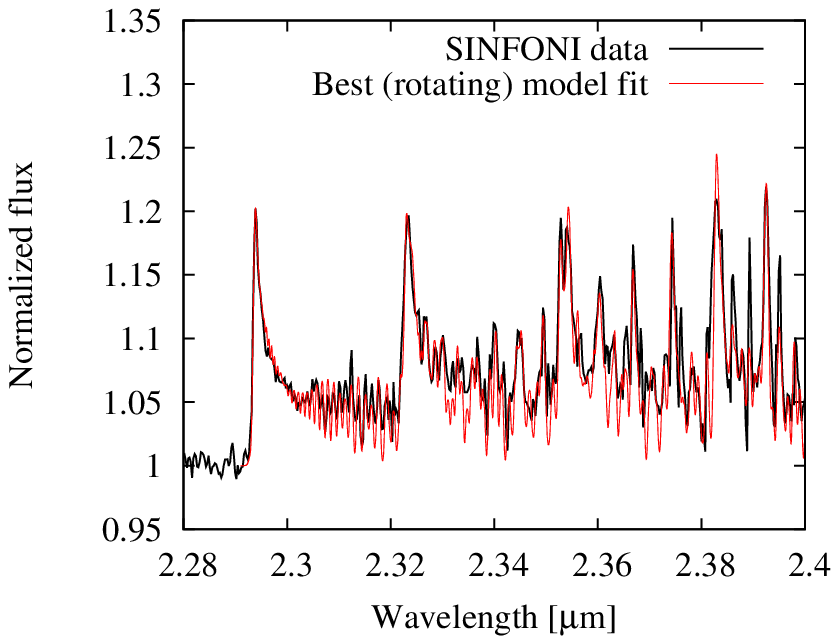}
  \includegraphics[angle=0,width=0.33\textwidth]{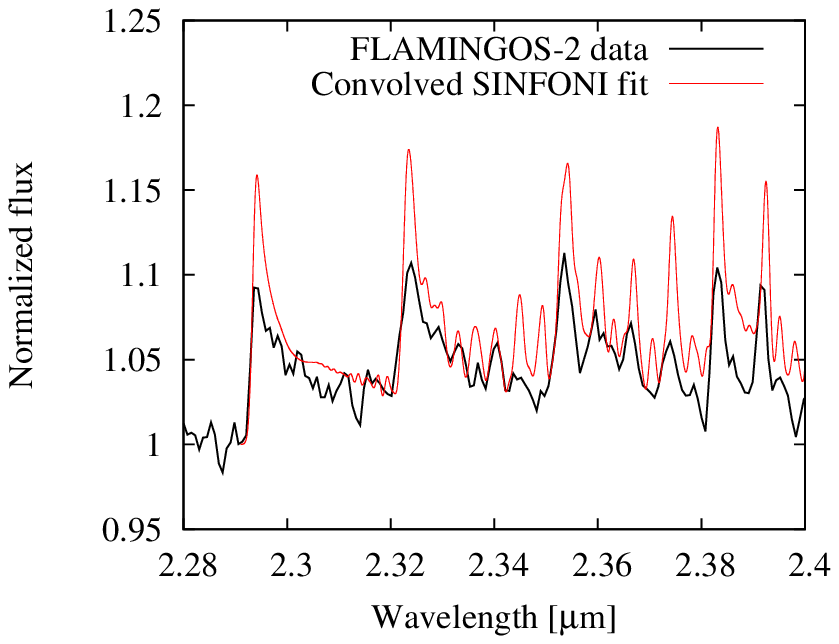}
  \includegraphics[angle=0,width=0.33\textwidth]{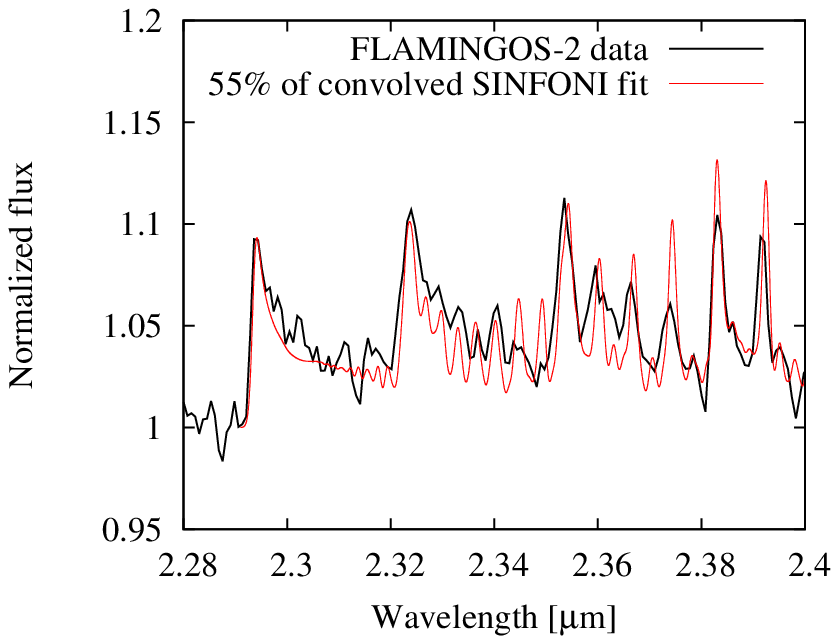}
  \caption{Left panel: Best model fit (in red) to the SINFONI CO band spectrum (in black), assuming a rotating CO ring scenario. Middle panel: Comparison between the FLAMINGOS-2 CO band spectrum (in black) and the SINFONI's best model fit (in red). The synthetic spectrum has been convolved to FLAMINGOS-2 resolution. Right panel: Comparison between the FLAMINGOS-2 CO band spectrum (in black) and the SINFONI's best model fit (in red), in which the emission was reduced to 55\% of its value in the SINFONI spectrum, while the Pfund emission was not altered.}
  \label{Figure-sinfoni-flamingos-co-model-fits}
  \end{centering}
\end{figure*}
%______________________________________________________________________________________

\subsection{The CO band head emission}\label{sec:co-bands}

The resolution of our near-IR spectra is too low to resolve the kinematics of the CO emitting region. However, to quantify the information of these data we proceeded in the following way. We considered the medium resolution SINFONI spectrum published by \citet{oks13}, in which a CO band head emission at 2.3 $\mu$m was detected for the first time. This observation corresponds to February 2012 and was interpreted using a non-rotating disc model. However, generally, it is assumed that the disc formation around B[e]SGs is related to fast rotation of the central stars. Thus, in order to improve \citeauthor{oks13}'s fit, we recomputed the physical parameters of the emitting CO gas assuming a rotating disc model. We applied the numerical codes developed by \citet{kra00} and \citet{kra09}. We considered that the CO emission arises from a thin ring in LTE in a rotating disc, where the ring temperature $T_{\mathrm{CO}}$, column density $N_{\mathrm{CO}}$ (along the line-of-sight), and rotational velocity $V_{\mathrm{rot}}$ projected to the line-of-sight are constant. We computed both the $^{12}$CO and $^{13}$CO emission spectra. Additionally, the hydrogen Pfund series emission spectrum was modelled using the code developed by \citet{kra00}. We assumed the Pfund lines are optically thin and in LTE, which is valid if the emitting material originates either in a wind or a shell. We computed the spectral lines using a pure Gaussian profile with a line width of 30 km s$^{-1}$ that then was convolved considering the SINFONI instrumental resolution.  
Figure \ref{Figure-sinfoni-flamingos-co-model-fits} (left panel) displays the best-fit model to the SINFONI first overtone spectrum, which was obtained with the following physical parameters: N$_{\mathrm{CO}}$ = 2.5$\pm$0.5 $\times$ 10$^{21}$ cm$^{-2}$, T$_{\mathrm{CO}}$ = 2\,500$\pm$200 K, $V_{\mathrm{rot}}$ = 35$\pm$5 km s$^{-1}$, a microturbulent velocity of $V_{\mathrm{turb}}$ = 5$\pm$1 km s$^{-1}$ and a ratio of \element[ ][12]{C}/\element[ ][13]{C} = 12.5$\pm$2.5. These parameters are similar to the ones obtained by \citet{oks13}, although the temperature value is a bit smaller. However, we should emphasize that from medium and low resolution data, we can only determine an upper limit for the Keplerian projected rotational velocity of the CO gas. This estimation is the maximum possible value that $V_{\mathrm{rot}}$ can reach in our computations before the theoretical spectrum begins to display features that are not observed in the SINFONI data. 

Then, we convolved the best-fit CO model to the resolution of the FLAMINGOS-2 spectrum and overplotted it to the FLAMINGOS-2 data which were taken almost two years after the acquisition of the SINFONI spectrum done by \citet{oks13}. This comparison is shown in the middle panel of Fig. \ref{Figure-sinfoni-flamingos-co-model-fits}. Obviously, this model overestimates the observations. However, if we reduce the CO emission to 55\% of its value in the SINFONI spectrum, without changing the Pfund emission, we reproduce the FLAMINGOS-2 CO band spectrum well enough (right panel). Therefore, we concluded that the intensity of the CO emission decreased in 2013, while the width of the band head seemingly did not change (at least not noticeable in our low-resolution spectra), which means that the rotational velocity of the emitting CO gas probably remained the same.

\subsection{Modelling optical forbidden emission lines}\label{sec:forbidden-lines}

One interesting property of the forbidden lines is that they are optically thin, hence their profiles carry the full kinematical information of their line-forming regions. Decoding this information from the [{\ion{Ca}{ii}}] $\lambda$\,7291 and [{\ion{O}{i}}] $\lambda\lambda$\,5577, 6300 line profiles is useful to trace the ionized and neutral atomic disc regions close to the star. Previous investigations have shown that both the CO emission bands and the forbidden lines of {\ion{Ca}{ii}} and {\ion{O}{i}} might share their forming regions, which can usually be described with a detached Keplerian rotating disc or ring model \citep{kra16, mar17}. Therefore, to constrain the kinematical properties of the forbidden line-forming regions in \object{LHA\,120-S\,35}, we applied a purely kinematical model \citep{are16a, kra16}, assuming that the emission originates from the same Keplerian rotating ring used in section \S \ref{sec:co-bands} to describe the CO emission bands. Thus, we computed a rotationally broadened line profile considering a projected rotational velocity $V_\mathrm{rot} \sim$ 35 km s$^{-1}$. We convolved this velocity considering a Gaussian component, $V_{\mathrm{gauss}}$, to model an extra broadening due to thermal and microturbulent motions of the gas (of a few km s$^{-1}$) and the instrument's spectral resolution. 

\begin{figure}
  \begin{centering}
  \includegraphics[angle=0,width=0.33\textwidth]{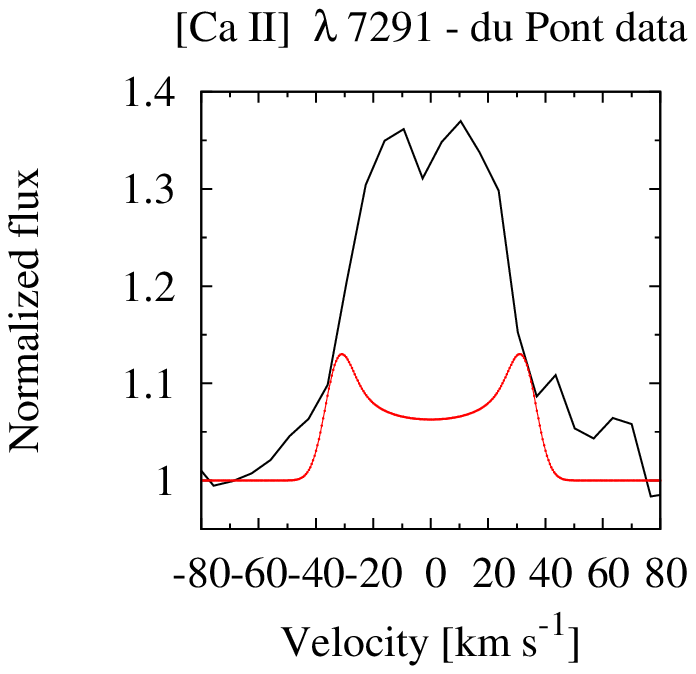}
  \includegraphics[angle=0,width=0.33\textwidth]{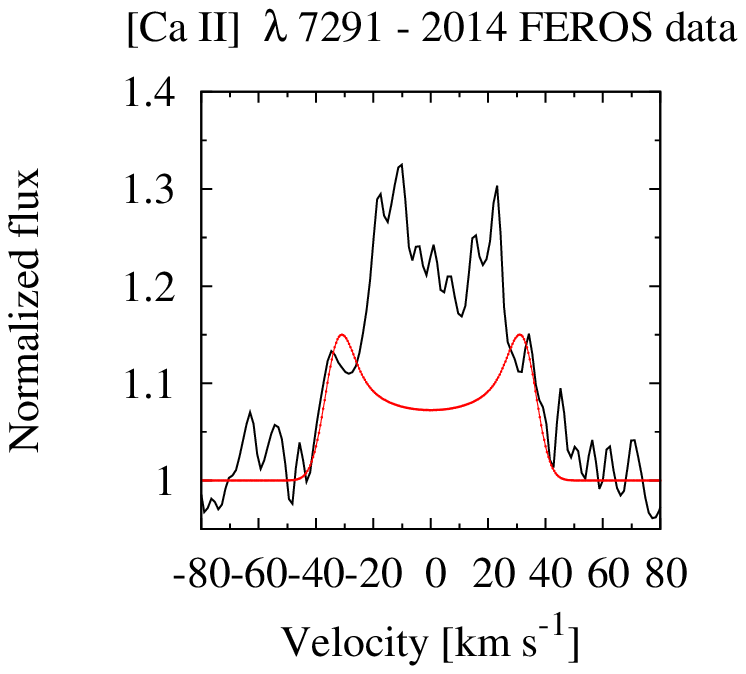}
  \caption{Comparison between the observed [{\ion{Ca}{ii}}] $\lambda$\,7291 line profile from the du Pont and 2014 FEROS spectra (top and bottom panels, respectively) and a computed theoretical profile considering a single ring model. Assuming the same rotational velocity as for the CO ring ($V_\mathrm{rot}$ = 35 km s$^{-1}$), the broadest part of the [{\ion{Ca}{ii}}] $\lambda$\,7291 line profile could be reproduced (data shown in black line, and the model in red).}
  \label{Figure-caii-bad-and-dupont-model-fits}
  \end{centering}
\end{figure}

If we compare the computed theoretical profile to the observed [{\ion{Ca}{ii}}] $\lambda$\,7291 line profiles from both the du Pont and FEROS spectra, we can see that the same (or a very similar) velocity component as for the CO ring seems to be present, as well as at least one more component that is required to fit the profiles properly (see Fig. \ref{Figure-caii-bad-and-dupont-model-fits}). Therefore, to reproduce simultaneously the FWHM of the observed lines, the separation between peaks and the wings, we considered a multi-ring model composed by several concentric rings, each ring characterized by the following two parameters: $V_\mathrm{rot}$ and $V_{\mathrm{gauss}}$.

\begin{figure*} 
  \begin{centering}
  \includegraphics[angle=0,width=0.33\textwidth]{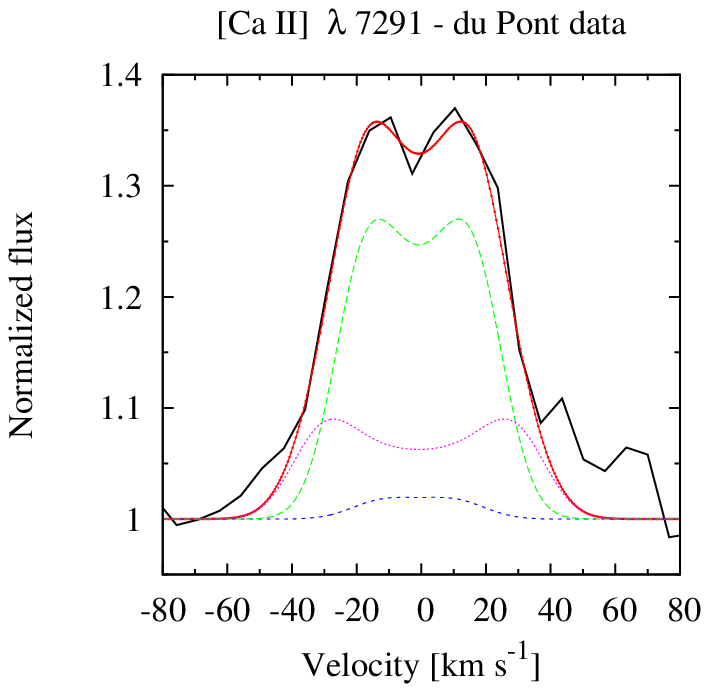}
  \includegraphics[angle=0,width=0.33\textwidth]{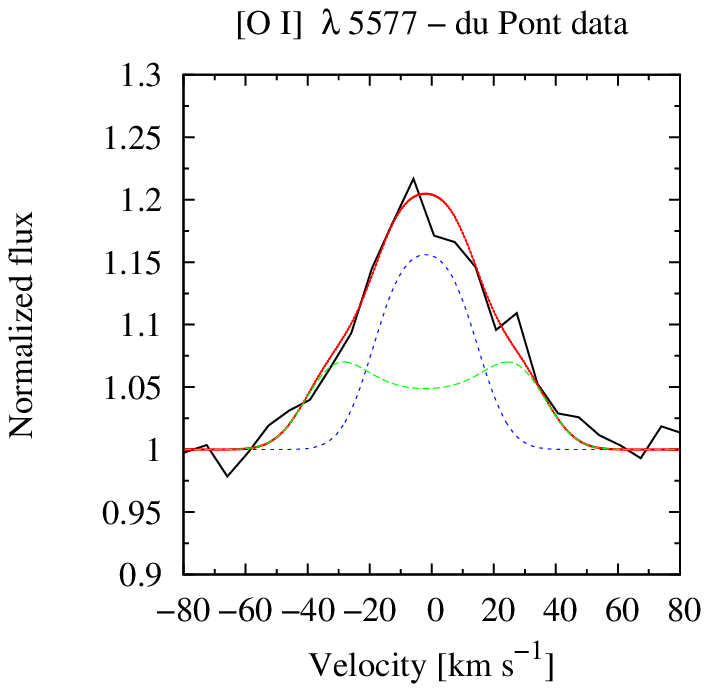} 
  \includegraphics[angle=0,width=0.33\textwidth]{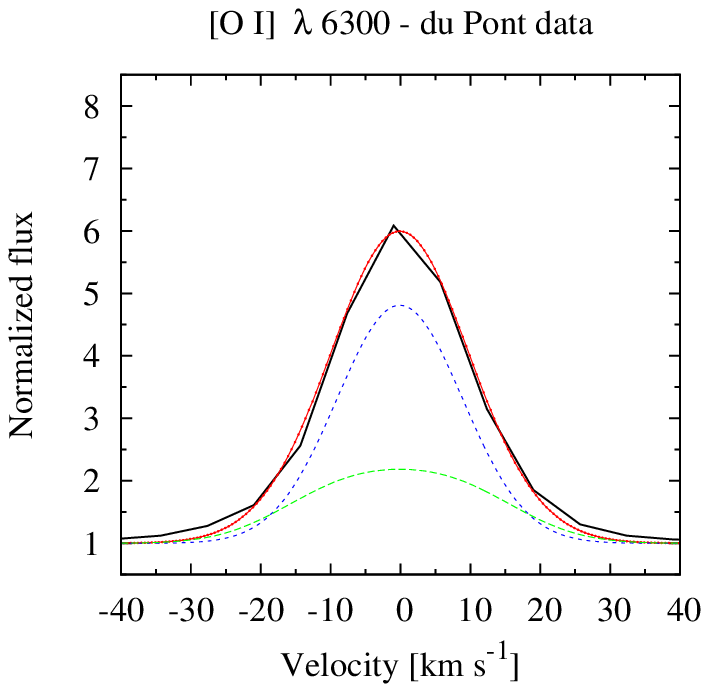}
  \caption{Model fits to the [{\ion{Ca}{ii}}] $\lambda$\,7291 (left panel), [{\ion{O}{i}}]  $\lambda$\,5577 (middle panel) and [{\ion{O}{i}}]  $\lambda$\,6300 (right panel) line profiles corresponding to the du Pont data. The computed profiles (in red lines) were obtained combining different emitting rings (in other colours) with different projected Keplerian rotational velocities (see Table \ref{Table-model-parameters}).}
  \label{Figure-du-Pont-model-fits}
  \end{centering}
\end{figure*}

A good fit was obtained to the [{\ion{Ca}{ii}}] $\lambda$\,7291 line profile for the du Pont spectra (see left panel of Fig. \ref{Figure-du-Pont-model-fits}), considering a model composed by three individual gas rings with different rotational velocities. The same multi-ring model also fitted well the [{\ion{O}{i}}] line profiles (see middle and right panels of Fig. \ref{Figure-du-Pont-model-fits}), considering two individual rings with the sets of parameters given in Table \ref{Table-model-parameters}, where we list $V_\mathrm{rot}$ and $V_\mathrm{gauss}$ in cols. 3 and 4, respectively.  

\begin{figure} 
  \begin{centering}
  \includegraphics[angle=0,width=0.33\textwidth]{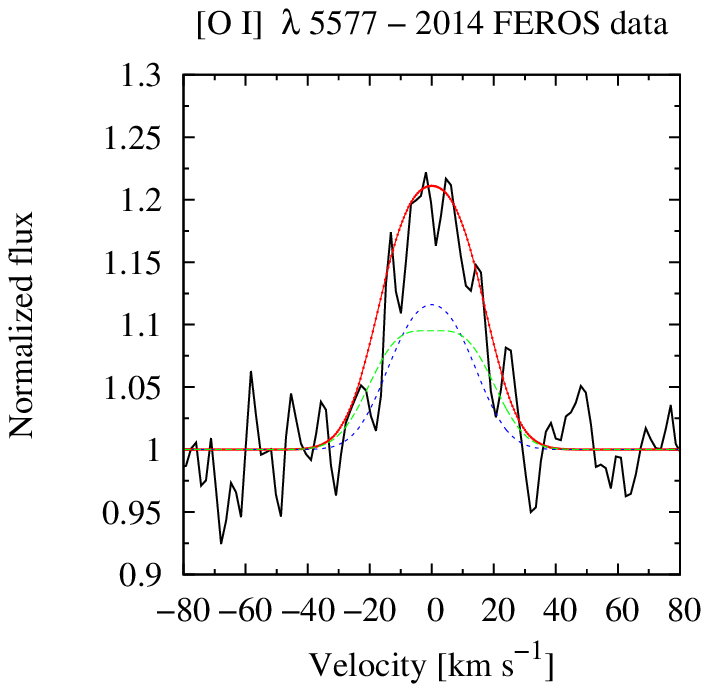} 
  \includegraphics[angle=0,width=0.33\textwidth]{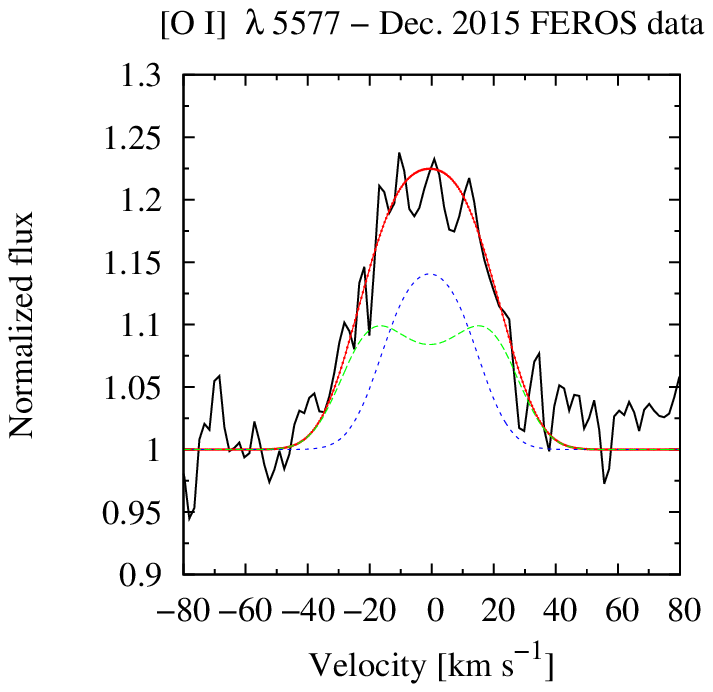}
  \caption{Model fits to the [{\ion{O}{i}}] $\lambda$\,5577 line profiles corresponding to the FEROS data (2014 and Dec 2015 data, top and bottom panels, respectively). The [{\ion{O}{i}}] $\lambda$\,5577 line from Nov 2015 FEROS spectra could not be modelled since it is very noisy.}
  \label{Figure-feros-oi-5577-model-fits}
  \end{centering}
\end{figure}

\begin{figure*} 
  \includegraphics[angle=0,width=0.33\textwidth]{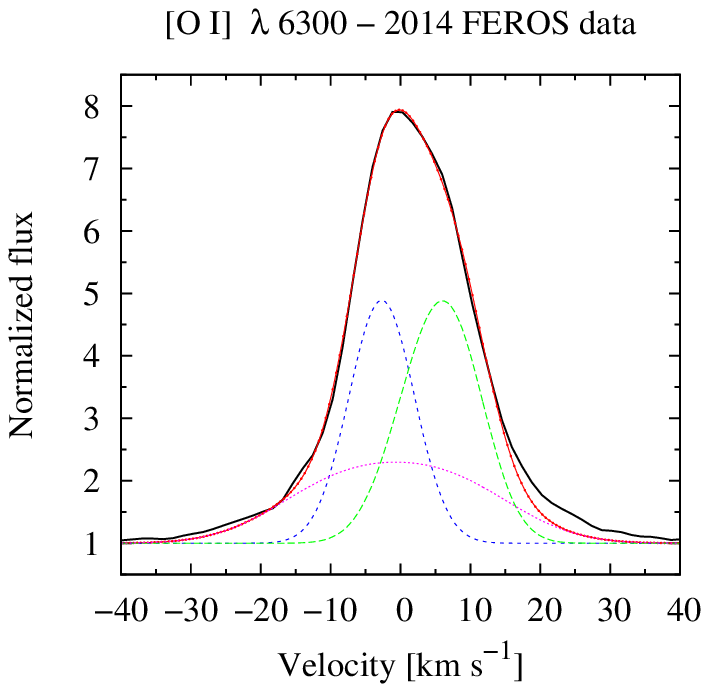}
  \includegraphics[angle=0,width=0.33\textwidth]{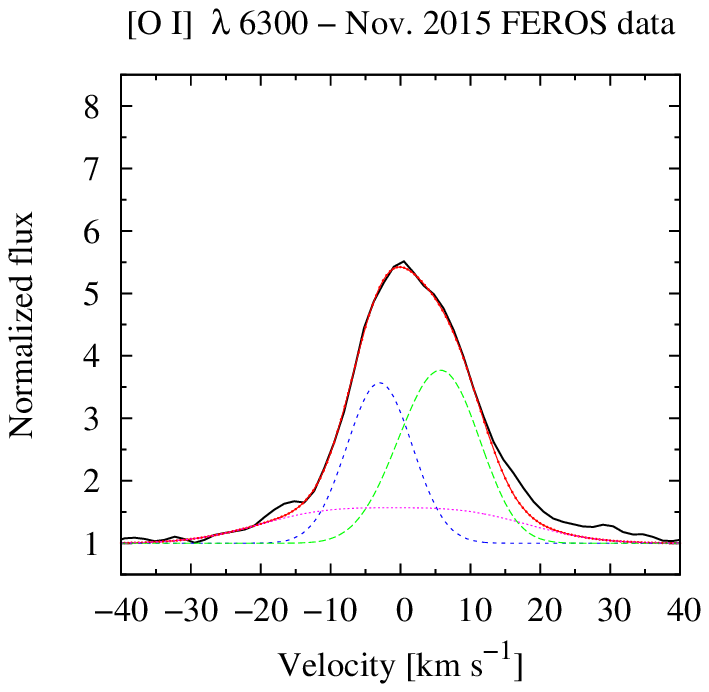}
  \includegraphics[angle=0,width=0.33\textwidth]{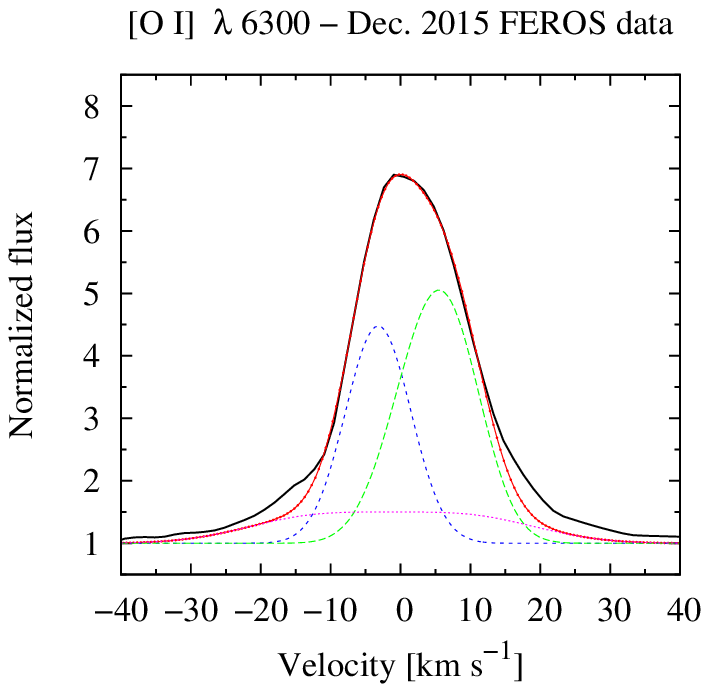}
  \caption{Model fits to the [{\ion{O}{i}}]  $\lambda$\,6300 line profiles corresponding to the FEROS spectra (2014, Nov 2015 and Dec 2015 data, left, middle and right panels, respectively). The computed profiles (in red line) were obtained combining different emission structures with different projected Keplerian rotational velocities (see Table \ref{Table-model-parameters}).}
  \label{Figure-feros-oi-6300-model-fits}
\end{figure*}

\begin{figure} 
  \begin{centering}
  \includegraphics[angle=0,width=0.33\textwidth]{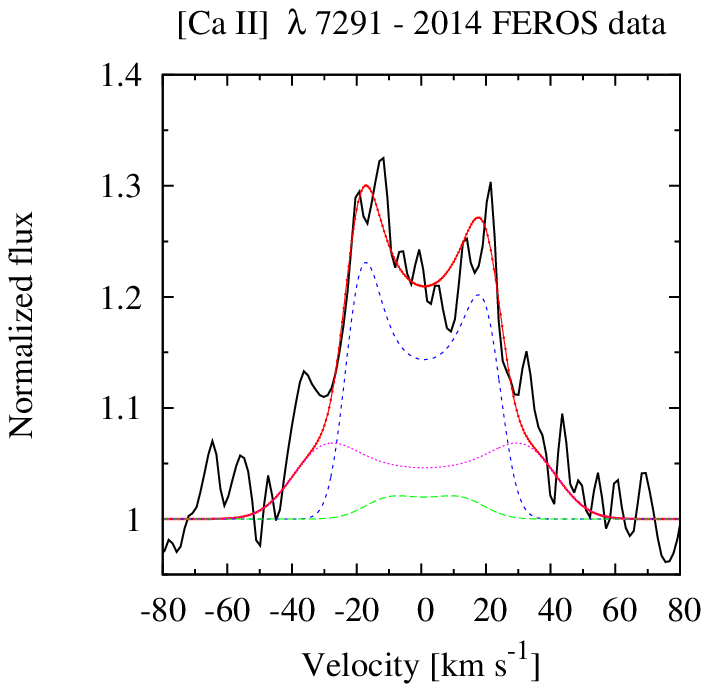}
  \includegraphics[angle=0,width=0.33\textwidth]{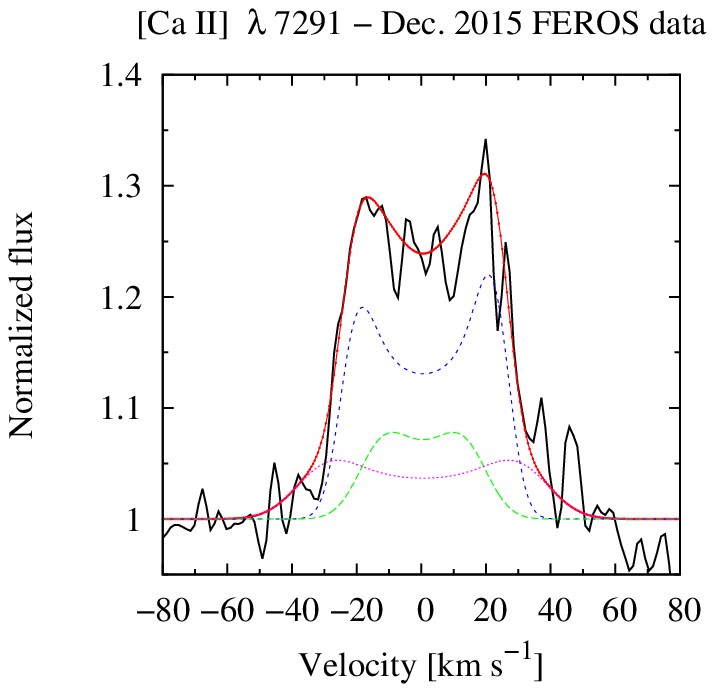}
  \caption{Model fits to the [{\ion{Ca}{ii}}] $\lambda$\,7291 profiles corresponding to the FEROS spectra (2014 and Dec 2015 data, top and bottom panels, respectively). The computed profiles (in red line) were obtained combining different emission structures with different projected Keplerian rotational velocities (see Table \ref{Table-model-parameters}). The [{\ion{Ca}{ii}}] $\lambda$\,7291 line from Nov 2015 FEROS data could not be modelled since it is very noisy.}
  \label{Figure-feros-caii-7291-model-fits}
  \end{centering}
\end{figure}

The [{\ion{O}{i}}] $\lambda$\,5577 line profile from the FEROS observations (see Fig. \ref{Figure-feros-oi-5577-model-fits}) was also well reproduced by a multi-ring model consisting of two rings, but with different sets of parameters than the ones obtained for the du Pont profile fits. However, the theoretical profile resulting from a multi-ring model did not match the observed [{\ion{Ca}{ii}}] $\lambda$\,7291 line profiles from FEROS data. Likewise, the [{\ion{O}{i}}] $\lambda$\,6300 line profiles could not be fitted considering a model with homogeneous concentric rings, even by varying the number of rings and/or their rotational velocities. To solve these discrepancies, it was necessary to consider partial ring structures to describe the emitting region components. Figures \ref{Figure-feros-oi-6300-model-fits} and \ref{Figure-feros-caii-7291-model-fits} show the good fits obtained for the [{\ion{O}{i}}] $\lambda$\,6300 and [{\ion{Ca}{ii}}] $\lambda$\,7291 line profiles, respectively. The best fitting parameters are given in Table \ref{Table-model-parameters}, where we also list the shape of the ring components (col. 5), either complete or partial, indicating in the last case the angle subtended by each ring segment, and the minimum and maximum projected rotational velocities used as integration limits of each ring component (cols. 6 and 7, respectively). 

We can summarize our results as follows. The gaseous equatorial disc where the forbidden emission lines of {\ion{O}{i}} and {\ion{Ca}{ii}} (and the molecular emission of CO) originate, shows regions of alternating density that form ring-like structures, either complete or fragmented. In some cases, the separation among these structures is not completely clear due to the large ring widths, since $V_{\mathrm{gauss}}$ may reach some km s$^{-1}$ beyond spectral resolution. We highlight four main emitting regions in Keplerian rotation as constituents of our model.
 
From the fitting to the du Pont spectra, we can identify four major emitting rings with projected rotational velocities spanning the ranges 36-32 km s$^{-1}$, 22-20 km s$^{-1}$, 17-11 km s$^{-1}$ and 3-1 km s$^{-1}$. The [{\ion{Ca}{ii}}] $\lambda$\,7291 emission line originates in the first three regions close to the star. The location of the [{\ion{O}{i}}] $\lambda$\,5577 line-forming regions coincides with the location of the first and third [{\ion{Ca}{ii}}] forming regions, while the [{\ion{O}{i}}] $\lambda$\,6300 line forms in the outermost two regions.

By fitting FEROS observations, we recognize a different disc spatial and kinematical configuration. The contribution of partial ring segments or fragmented clumps in Keplerian rotating regions is revealed. However, on comparing the modelling for both set of observations, we have to consider that the du Pont data were binned and thus have a lower resolution than the FEROS data. This may yield differences in the line profile shapes and thus different modelling results. Nevertheless, the possibility of a real change in the ring's structure might not be discarded.

The 2014 forbidden emission lines indicate the presence of four distinct emitting regions with projected rotational velocities of $\sim$ 37-35 km s$^{-1}$, $\sim$ 23-21 km s$^{-1}$, $\sim$ 18-9 km s$^{-1}$ and $\sim$ 3-1 km s$^{-1}$. Our results suggest that the inner forming-region of [{\ion{Ca}{ii}}] line is a complete ring, while the second one seems to be a ring segment. The third emitting region, where both {\ion{Ca}{ii}} and {\ion{O}{i}} coexist, seems to present clumped structures. Finally, the outermost region, where the {\ion{O}{i}} emission contributes to the forbidden transition at $\lambda$\,6300 \AA, looks like a fragmented ring segment. 

For the 2015 FEROS data, we obtain more or less the same distinct multi-ring structure, although for the November spectrum we only modelled the [{\ion{O}{i}}] $\lambda$\,6300 emission line, since the [{\ion{O}{i}}] $\lambda$\,5577 and the [{\ion{Ca}{ii}}] $\lambda$\,7291 lines are distorted by the noise. We distinguish the first ($\sim$ 37-35 km s$^{-1}$) and third ($\sim$ 17-9 km s$^{-1}$) regions observed in 2014. The second ($\sim$ 26-22 km s$^{-1}$) looks like an inhomogeneous ring with the coexistence of {\ion{Ca}{ii}} and {\ion{O}{i}}, which contributes to the [{\ion{O}{i}}] $\lambda$\,5577 emission. The fourth outermost partial ring ($\sim$ 3-1 km s$^{-1}$) gives rise, as before, to the [{\ion{O}{i}}] $\lambda$\,6300 line.

The good matches between the computed profiles and the observed data showed a trend with respect to the multi-ring configuration: while the outermost rings are partial, the innermost are complete. It seems that the rings fragment or dissolve as they are getting older.

%______________________________________________________________________________________
% TABLE 1------------------------------
\begin{table*}[bpt]
  \begin{center}
  \caption{Best-fit model parameters to the forbidden line profiles for \object{LHA\,120-S\,35}.} \label{Table-model-parameters}
  \begin{tabular}{llcccccc}
\hline
\\
  Spectra            & Line                                     & $V_\mathrm{rot}$        & $V_\mathrm{gauss}$        & Ring structure      & $V_\mathrm{rot}^\mathrm{min}$  & $V_\mathrm{rot}^\mathrm{max}$\\ 
                     &                                          & $[\mathrm{km\,s^{-1}}]$ & $[\mathrm{km\,s^{-1}}]$  &                     & $[\mathrm{km\,s^{-1}}]$       & $[\mathrm{km\,s^{-1}}]$\\
\hline
2008 du Pont         &                                          &                        &                          &                     &                               &                       \\
                     & $\mathrm{[O\, I]\, \lambda}$ 5577        & 34.0$\pm$2             & 12.5$\pm$2               & complete            & -34.00                        & 34.00                 \\
                     &                                          & 13.0$\pm$2             & 12.5$\pm$2               & complete            & -13.00                        & 13.00                 \\
\cdashline{2-8}[.4pt/2pt]
\\
                     & $\mathrm{[O\, I]\, \lambda}$ 6300        & 12.0$\pm$1             & 12.5$\pm$1               & complete            & -12.00                        & 12.00             &   \\
                     &                                          &  2.0$\pm$1             & 12.5$\pm$1               & complete            & -2.00                         & 2.00                  \\
\cdashline{2-8}[.4pt/2pt]
\\
                     & $\mathrm{[Ca\, II]\, \lambda}$ 7291      & 34.0$\pm$1             & 12.5$\pm$1               & complete            & -34.00                        & 34.00                 \\
                     &                                          & 21.0$\pm$1             & 12.5$\pm$1               & complete            & -21.00                        & 21.00                 \\
                     &                                          & 16.0$\pm$1             & 12.0$\pm$1               & complete            & -16.00                        & 16.00                 \\
\hline
2014 FEROS           &                                          &                        &                          &                     &                               &                       \\
                     & $\mathrm{[O\, I]\, \lambda}$ 5577        & 16.0$\pm$2             & 12.5$\pm$2               & complete            & -16.00                       & 16.00                 \\   
                     &                                          & 11.0$\pm$2             & 12.5$\pm$2               & complete            & -11.00                       & 11.00                 \\
\cdashline{2-8}[.4pt/2pt]
\\
                     & $\mathrm{[O\, I]\, \lambda}$ 6300        & 12.0$\pm$1             & 12.5$\pm$1               & complete            & -12.00                       & 12.00                 \\
                     &                                          & 10.0$\pm$1             &  6.5$\pm$1               & partial (90$^\circ$) & 0.00                        & 10.00                 \\
                     &                                          &  2.0$\pm$1             &  6.5$\pm$1               & partial (10$^\circ$) & -2.00                       & -1.96                 \\ 
\cdashline{2-8}[.4pt/2pt]
\\
                     & $\mathrm{[Ca\, II]\, \lambda}$ 7291      & 36.0$\pm$1             &  12.5$\pm$1              & complete            & -36.00                       & 36.00                 \\
                     &                                          & 22.0$\pm$1             &  6.5$\pm$1               & partial (170$^\circ$)& -22.00                      & 21.66                 \\
                     &                                          & 15.0$\pm$1             &  9.5$\pm$1               & complete            & -15.00                       & 15.00                 \\
\hline
Nov 2015 FEROS      &                                          &                        &                          &                     &                              &                       \\
                     & $\mathrm{[O\, I]\, \lambda}$ 5577        &  ...                   &  ...                     & ...                 &  ...                         & ...                   \\
\cdashline{2-8}[.4pt/2pt]
\\
                     & $\mathrm{[O\, I]\, \lambda}$ 6300        & 15.0$\pm$1             & 12.5$\pm$1               & complete            & -15.00                      & -15.00                 \\
                     &                                          & 10.0$\pm$1             &  6.5$\pm$1               & partial (90$^\circ$) & 0.00                       &  10.00                 \\
                     &                                          &  2.0$\pm$1             &  6.5$\pm$1               & partial (10$^\circ$) & -2.00                      & -1.96                  \\
\cdashline{2-8}[.4pt/2pt]
\\
                     & $\mathrm{[Ca\, II]\, \lambda}$ 7291      &  ...                   &  ...                     & ...                 & ...                         & ...                    \\
\hline
Dec 2015 FEROS      &                                          &                        &                          &                     &                             &                       \\
                     & $\mathrm{[O\, I]\, \lambda}$ 5577        & 24.0$\pm$2             & 12.5$\pm$2               & complete            & -24.00                      & 24.00                 \\
                     &                                          & 14.0$\pm$2             & 12.5$\pm$2               & complete            & -14.00                      & 14.00                 \\ 
\cdashline{2-8}[.4pt/2pt]
\\
                     & $\mathrm{[O\, I]\, \lambda}$ 6300        & 16.0$\pm$1             & 12.5$\pm$1               & complete            & -16.00                      & 16.00                 \\
                     &                                          & 10.0$\pm$1             &  6.5$\pm$1               & partial (90$^\circ$) & 0.00                       & 10.00                 \\
                     &                                          &  2.0$\pm$1             &  6.5$\pm$1               & partial (10$^\circ$) & -2.00                      & -1.96                 \\
\cdashline{2-8}[.4pt/2pt]
\\
                     & $\mathrm{[Ca\, II]\, \lambda}$ 7291      & 34.0$\pm$1             &  12.5$\pm$1              & complete            & -34.00                      & 34.00                 \\
                     &                                          & 24.0$\pm$1             &  6.5$\pm$1               & partial (170$^\circ$)& -23.63                     & 24.00                 \\
                     &                                          & 16.0$\pm$1             &  9.5$\pm$1               & complete             & -16.00                     & 16.00                 \\
\hline
2012 SINFONI         &                                          &                        &                          &                      &                            &                       \\
                     & CO bands                                 & $\leq$ 35.0$\pm$5      & 5.0$\pm$1                & complete             & -35.00                     & 35.00                 \\
\hline
  \end{tabular}
  \tablefoot{The last line shows the best-fit model parameters to the CO emission bands.}
  \end{center}
\end{table*}
% -------------------------------------
%______________________________________________________________________________________
%______________________________________________________________________________________
% FIGURE 14-----------------------------
% -------------------------------------
%______________________________________________________________________________________

%______________________________________________________________________________________

\section{Discussion and conclusions}\label{sec:discussion}

\begin{figure}
  \begin{centering}
  \includegraphics[angle=0,scale=0.7]{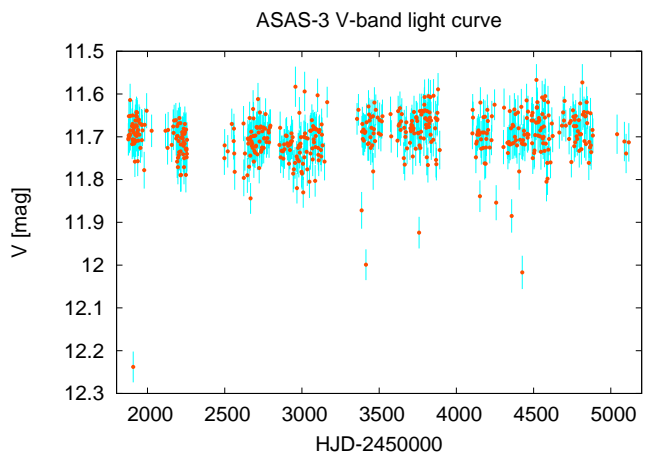}
  \caption{ASAS-3 $V$-band light curve of \object{LHA\,120-S\,35}. Although no reliable period was found to describe the variations, a sort of long period (P > 1750 d) oscillation may be present on which there may be overlapping irregular light variations with amplitudes of about 0.2 mag.}
  \label{Figure-asas-lightcurve}
  \end{centering}
\end{figure}

\object{LHA\,120-S\,35} displays not only spectral variations but also photometric variability. Its $V$-band light curve, downloaded from the ASAS-3 Photometric $V$-band Catalogue \citep{poj02}, is displayed in Fig. \ref{Figure-asas-lightcurve}. The data span a time interval of nine years, from November 2000 to November 2009. We checked the quality of each data point and kept only those that were labelled as quality A. The final number of reliable measurements were thus reduced to 517 values. These are shown in Fig. \ref{Figure-asas-lightcurve} with their corresponding errorbars. We performed a period analysis with the AOV Periodogram Routines \citep{sch89, sch96, sch06} and we did not find any reliable period to describe the variations. This could be due to the contamination of the photometry by a nearby star separated less than 12 arcsec, which is not resolved by the ASAS aperture photometry. However, a sort of long period (P > 1750 d) oscillation may be present in the $V$-band light curve of \object{LHA\,120-S\,35} on which there may be overlapping irregular light variations with amplitudes of about 0.2 mag, reaching up to $\sim$0.7 mag in some sporadic points. Traditionally, it has been considered that B[e]SGs show little or no photometric variations \citep{zic86, lam98}. However, the increase in the number of photometric observations not only has enlarged the observed B[e]SG sample but also has improved the temporal distribution of the data. From this, it was possible to find that the photometric behaviour of some B[e]SGs disagrees with the historical picture as, for example, in \object{LHA\,115-S\,18} which is highly variable on multiple timescales with amplitudes ranging from $\sim$0.1 to 1.0 mag \citep{gen02,cla13a,mar14}, the B[e]SG/X-ray binary \object{CI\,Cam} which showed evidence of variability both in the X-ray outburst and in quiescence \citep{cla00}, \object{VFTS\,698} with photometric variations of $\sim$0.6 mag on timescales ranging from days to years \citep{dun12} and \object{MWC\,349A} with periodic variations of $\sim$0.6 mag in amplitude \citep{yud96,jor00}. It is important to remark that all these stars are proposed or considered as members of binary systems \citep{cla13a,bar13,kim12,sal17} while the binary nature of \object{LHA\,120-S\,35} is still inconclusive. 

Spectroscopic variations are observed in both the optical and infrared wavelength ranges. Basically all optical lines display variabilities in shape and/or strength. The most pronounced changes are seen in the forbidden and permitted \ion{O}{i} lines, these lines being more intense in 2014. Similarly, the [\ion{Fe}{ii}] lines show strong variability with the 2014 spectra displaying the highest intensity.

The spectroscopic analysis of \object{LHA\,120-S\,35} suggests that the star has a strong and variable bipolar clumped wind, which is revealed throughout the P-Cygni line profiles of \ion{H}{i}, \ion{Fe}{ii} and \ion{O}{i}. Density enhancements moving across the wind are seen superimposed to the absorption P-Cygni profiles (NACs) in the first Balmer lines at different velocities, showing variations in both intensity and velocity. NACs features in the optical lines were also seen in the luminous blue variables (LBVs). \citet{gro11} modelled the moving NACs seen in the LBV \object{AG\,Car} and explained their behaviours due to abrupt changes in the wind-terminal velocity, associated to the bi-stability jump. This jump is related to a change in the degree of ionization in the wind \citep{vin00} which is caused by the recombination of \ion{Fe}{iv} to \ion{Fe}{iii}. However, observations of early B SGs showed that there is a gradual decrease in the wind terminal velocities as a function of effective temperature, instead of a 'jump' \citep{cro06}, and that the mechanism of bi-stability could be present among stars with temperatures ranging between 18\,000 K and 23\,000 K \citep{mar08}. The effective temperature of \object{LHA\,120-S\,35} \citep[22\,000 K,][]{gum95} falls into this range, so a similar phenomenology could take place in this star.
%______________________________________________________________________________________
%FIGURE 15-----------------------------
% -------------------------------------
%______________________________________________________________________________________

Additionally, the double-peaked features seen in the emission component of the Balmer P-Cygni profiles with the blue-shifted central absorption suggest the presence of a slow outflowing rotating circumstellar disc. A similar global structure composed by an equatorial disc perpendicular to a bipolar flow has been proposed to interpret the observations at millimeter and radio wavelengths for the B[e]SGs \object{MWC\,349} \citep{dan01,taf04} and \object{Wd1-9} \citep{fen17} and the optical and infrared observations of MWC 137 \citep{kra17}.

The shape of the Paschen lines seems to indicate a two-component forming region as well, a weak but broad, maybe double-peaked profile underneath a narrower single-peaked profile, where the first might come from the disc or rings and the second from the wind. Considering that intensity changes are only observed in the lower members of the Paschen series, which are formed in much outer regions than the higher members, they might be influenced by the clumpy structure of the rings.

In the outer part of this disc, we found spectroscopic evidences of detached multi-ring structures with density variations along the disc. The inner ring has a sharp edge where [\ion{Ca}{ii}] and [\ion{O}{i}] lines share their forming region with the CO molecular bands. The outermost regions show a complex structure, with fragmented clumps and/or partial-ring features traced by \ion{Ca}{ii} and \ion{O}{i}. The shape of these complete and partial extended ring structures together with their dynamics (apparent expansion, fragmentation, re-formation at high velocities within short time scales) reminds of (spiral) density waves that can be excited in circumstellar discs via gravitational forces and which have been observed in planetary systems \citetext{e.g. \citealt{shu16} and references therein} and in protoplanetary discs \citetext{e.g. \citealt{fuk04}; \citealt{per16}}. However, our observational data set is too sparse to confirm such a scenario and to identify its possible mechanism amongst which are the interaction with a companion, planet-disc interaction, or gravitational instabilities.

%______________________________________________________________________________________
%FIGURE 16-----------------------------
% -------------------------------------
%______________________________________________________________________________________

%______________________________________________________________________________________
%FIGURE 17-----------------------------
% -------------------------------------
%______________________________________________________________________________________

In order to interpret the spectroscopic variations of the forbidden lines, our model computations were performed assuming Keplerian (complete or partial) ring-like structures located in the equatorial plane. However, this scenario is not unique. The observed ring segments could be the result of the projection onto the sky plane of a three-dimensional structure (for example, an hourglass nebula) with density enhancements. If we consider that \object{LHA\,120-S\,35} is an evolved object displaying a circumstellar environment that is oxygen-rich and contains a large amount of dust, the formation of such an hourglass-like structure might result from wind-wind interactions of the current wind of the star in its presumable post-red supergiant (post-RSG) state with previously ejected material from the RSG \citep{chi08}. However, in this case an hourglass-like structure can only be achieved if the wind is not spherically symmetric anymore.

%______________________________________________________________________________________
%FIGURE 18-----------------------------
% --------------------------------------
%______________________________________________________________________________________

Interestingly, \citet{dav07} suggested that the direct successors of RSGs, the yellow hypergiants (YHGs), change their mass-loss behaviour from spherical to axisymmetric during their passage through the instability domain called Yellow Void. This change results in the formation of an equatorial disc and a bipolar wind, and it is assumed that YHGs appear as B[e]SGs after they reach the blue edge of the instability domain \citep{dav07,are17}. The YHGs pulsations are believed to trigger the enhanced mass-loss \citep{dej98}. Whether such a scenario also holds for the B[e]SGs cannot be firmly said. However, \citet{kra16} recently postulated that the B[e]SG \object{LHA\,120-S\,73} displays indications for pulsation activity based on the observed \ion{He}{i} line profile variability. Inspection of the profiles of the \ion{He}{i} $\lambda$\,5876 line, which seems to be the only unblended absorption line in the optical spectrum of \object{LHA\,120-S\,35} reveals that this line also shows variability in its profile that might originate from stellar pulsations.  

Alternatively, even though evidence indicative of binarity has not yet been found in \object{LHA\,120-S\,35}, a Roche lobe overflow phase in an interacting binary or a colliding-wind binary system could also be suitable scenarios to analyse the complex circumstellar ring-like (and possibly spiral) structures of \object{LHA\,120-S\,35}, as for example in \object{RY\,Scuti} \citep{smi11} or \object{Wd1-9} \citep{cla13b}. However, emission in key lines attributed to binarity as \ion{N}{iii} $\lambda \lambda$\,4634--42 and \ion{He}{ii} $\lambda$\,4686 is not seen, conversely to what was detected in B[e]SGs known to be binaries like \object{LHA 115-S 18} \citep{cla13a} and \object{HD 38489} \citep{mas14}. The \ion{He}{ii} $\lambda$\,4686 line is highly variable in \object{LHA 115-S 18}, and it is relevant to highlight that it may even disappear at certain times \citep{tor12}. Furthermore, the orbital motion in a binary system could also cause variations in the line-profiles. Unfortunately, the data gathered in this work is not enough to make a deep analysis.

%______________________________________________________________________________________
% FIGURE 19-----------------------------
% --------------------------------------
%______________________________________________________________________________________
%______________________________________________________________________________________
% FIGURE 20-----------------------------
%Wide-field Infrared Survey Explorer
\begin{figure*}[!hbt]
  \begin{centering}
  \includegraphics[angle=0,width=9.80 cm]{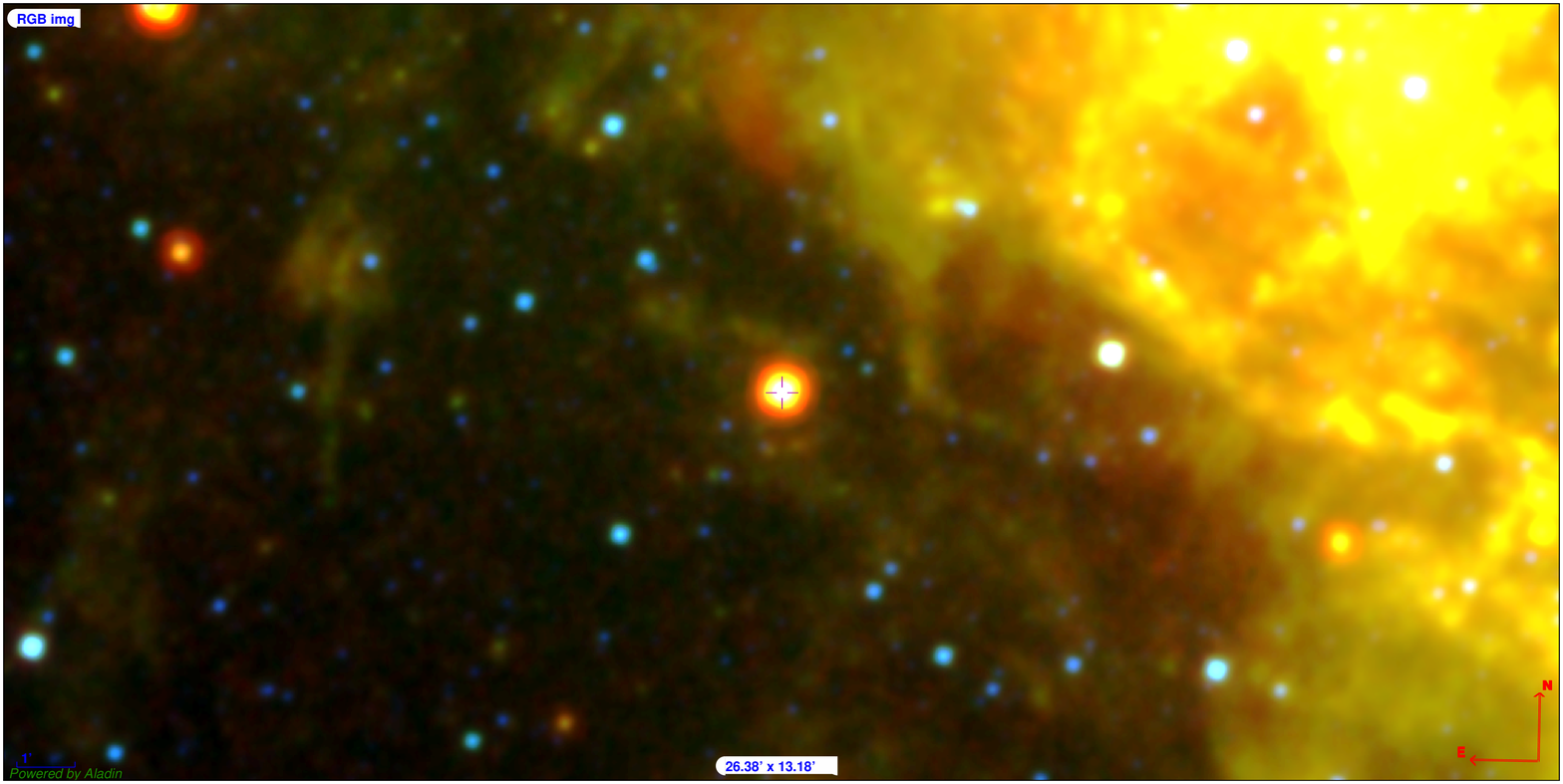}
  \includegraphics[angle=0,width=9.80 cm]{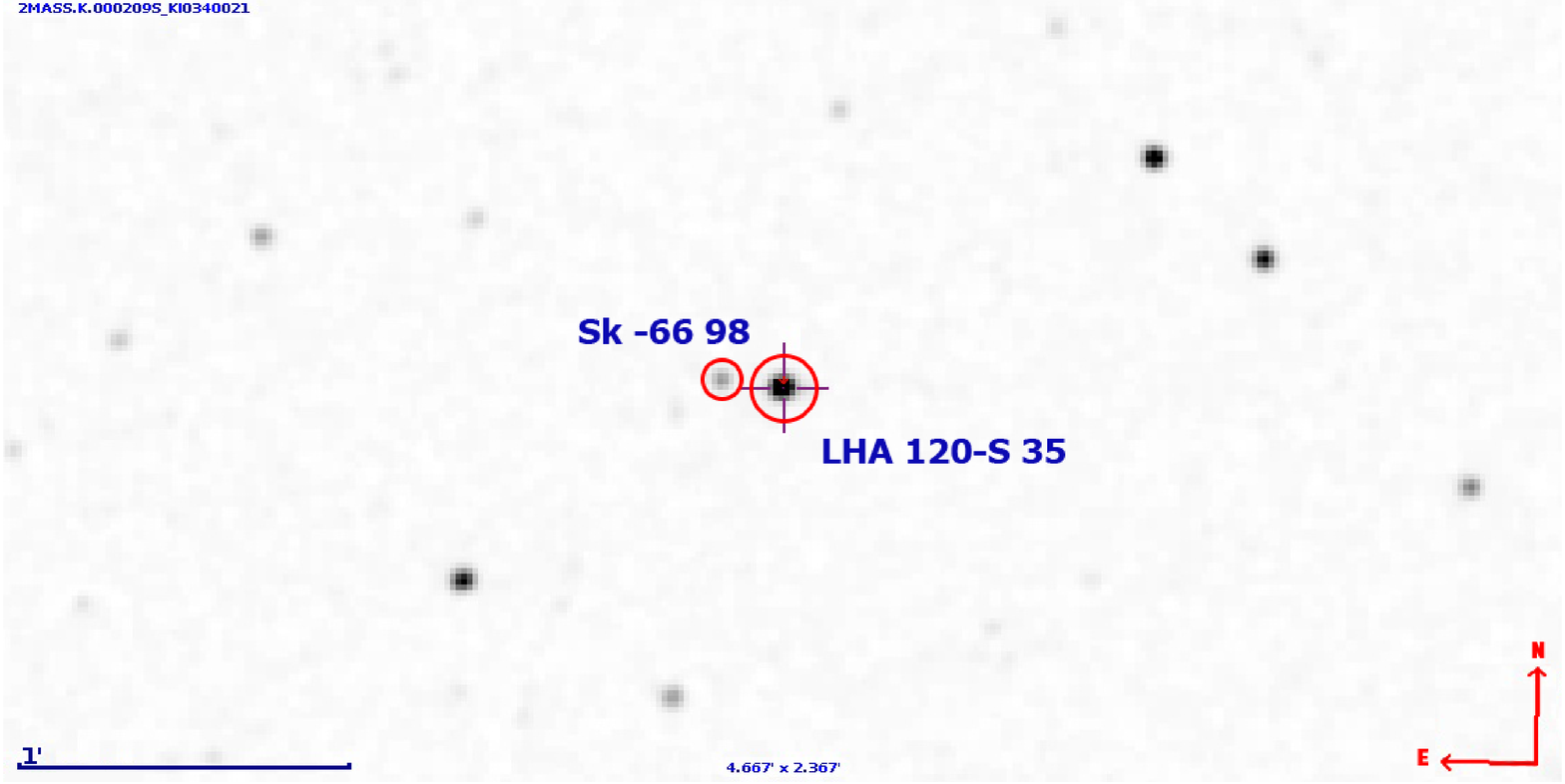}
  \caption{Top panel: Representative colour RGB WISE image ($\sim$26$\times$13\,arcmin$^2$) of \object{LHA\,120-S\,35} and its surroundings made from images at 4.6 $\mu$m (in blue), 12 $\mu$m (in green) and 22 $\mu$m (in red) with an angular resolution of 6.4, 6.5 and 12.0 arcsec, respectively. Bottom panel: 2MASS $K$-band image ($\sim$4.7$\times$2.4\,arcmin$^2$) showing \object{LHA\,120-S\,35} (indicated by a cross) along with the neighbouring star \object{Sk\,-66\,98}, which remains unresolved in the WISE image. Both images are orientated north up and east to the left.}   
  \label{Figure-s35-wise}
  \end{centering}
\end{figure*}
% -------------------------------------
%______________________________________________________________________________________

\begin{figure}[!bh]
  \begin{centering}
  \includegraphics[angle=0,width=0.5\textwidth]{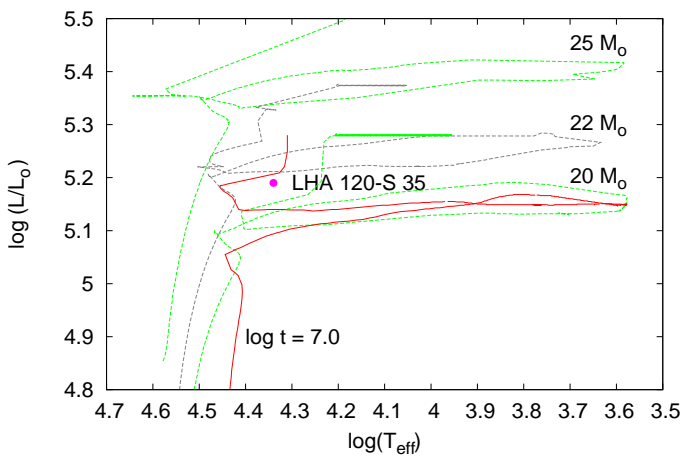}
  \caption{Hertzsprung-Russell diagram showing the position of \object{LHA\,120-S\,35}. The evolutionary tracks for a rotating massive star with 20 $M_{\odot}$ and 25 $M_{\odot}$ taken from \citet{eks12} are plotted in green, from which we obtained by interpolation the evolutionary track for a star with 22 $M_{\odot}$ (in grey). These tracks correspond to solar metallicity $Z$=0.014. An isochrone of $\log\,t$ = 7.0 is shown in red.} 
  \label{Figure-s35-HR-diagram}
  \end{centering}
\end{figure}

Therefore, to unveil the proper scenario for the circumstellar material distribution, high spatial resolution images would be required. However, this is a difficult task, especially for stars located outside the Milky Way, and currently available images from the WISE (Wide-field Infrared Survey Explorer) and Spitzer space telescopes are of too low spatial resolution to provide reasonable information about the environment of \object{LHA\,120-S\,35}. Figure \ref{Figure-s35-wise} (top panel) shows a WISE image of a field containing \object{LHA\,120-S\,35}, composed by observations obtained with three of the four infrared detectors aboard the satellite \citep{wri10}. The composite image shows in blue the infrared light at 4.6 $\mu$m (W2 band), which is mainly from background stars. The light at 12 $\mu$m (W3 band) and 22 $\mu$m (W4 band), arising primarily from warm dust, is represented by green and red, respectively. Figure \ref{Figure-s35-wise} (bottom panel) shows a 2MASS $K$-band image, revealing the presence of a neighbouring star of \object{LHA\,120-S\,35} at $\sim$12 arcsec separation, \object{Sk\,-66\,98}, which is unresolved on the WISE image. \object{LHA\,120-S\,35} has been identified with the infrared source \object{SL\,482-IR1} that was classified by \citet{loo05} as a hot and dusty object with a $T_{\rm{eff}}$ = 20\,000 K and $\log\,(L/L_{\odot})$ = 5.19. These authors consider \object{SL\,482-IR1} as a member of the young stellar cluster \object{SL\,482} (also named as \object{LMC\,1015}) with a cluster age of $\log\,t$ = 7.2$\pm$0.24 \citep{loo05}. Other authors estimated different values for the cluster age ranging from $\log\,t$ = 7.6 to 8.0 \citep{gla10,ko13,bit17}. Comparing the position of \object{LHA\,120-S\,35} in the Hertzsprung-Russell diagram to the evolutionary track of a massive star with $\sim$22 $M_{\odot}$ \citep{eks12}, we corroborated that an isochrone of $\log\,t$ = 7.0 can reconcile the cluster age derived by \citet{loo05} with the position of a post-RSG object (see Fig. \ref{Figure-s35-HR-diagram}). However, this result needs to be taken with caution, since the tracks computed by \citet{eks12} consider solar metallicity. So, for a low metallicity LMC star this scenario can be different.
On the other hand, \citet{loo05} also reported that \object{SL\,482-IR1} has IR colours similar to other low-excitation planetary nebulae, and they modelled the IR spectral energy distribution assuming an oxygen-rich silicate dust shell envelope with a T$_{\rm{dust}}$ = 750 K, which might be detached. However, both the B[e]SGs and the planetary nebulae are in the same locus on the $J-H/H-K_{\rm s}$ colour-colour diagram \citep{oks13, rei14}, which makes it difficult to assign a classification to the star based on its infrared colours. But, as the luminosity of \object{LHA\,120-S\,35} is higher than the one expected for low-excitation planetary nebulae in the MCs \citep{kal91}, we can discard its classification as a planetary nebula. 
Therefore, if we consider the classification of \object{SL\,482-IR1} as a B[e]SG and the age inferred by \citet{loo05}, \object{LHA\,120-S\,35} would be the third B[e]SG known up to now belonging to a young stellar cluster. Recently, the \object{LHA\,120-S\,35} membership to the cluster \object{SL\,482} has been discussed by \citet{bit17}, who were able to determine the members of the star cluster using a robust decontamination technique, providing a probability of more than 92$\%$ that \object{LHA\,120-S\,35} is member of the cluster. The previously reported B[e]SGs associated to open clusters are: \object{LHA\,120-S\,111}, which belongs to the compact cluster \object{NGC\,1994} of the LMC  \citep{mel01,lor88} and \object{Wd1-9}, which is a member of the massive Galactic stellar cluster \object{Westerlund 1} \citep{cla13b, fen17}. The small number of B[e]SGs found in stellar clusters is reasonable taking into account that their lifetime in this post-main-sequence phase is relatively short.

Finally, to determine the proper rotation velocity of the different gas tracers and to study the evolution of the inhomogeneities along the circumstellar structure as well as the origin of the variability of the absorption line-profiles,  high-resolution observations regularly taken are needed.

%______________________________________________________________________________________

%______________________________________________________________________________________

\begin{acknowledgements}

We thank the referee Dr. Simon Clark for his valuable comments that helped to improve the manuscript. This publication makes use of the NASA Astrophysics Data System (ADS), the SIMBAD database, operated at CDS, Strasbourg, France, and the data products from the Wide-field Infrared Survey Explorer, which is a joint project of the University of California, Los Angeles, and the Jet Propulsion Laboratory/California Institute of Technology, funded by the National Aeronautics and Space Administration. We would like to thank the observers (S. Ehlerova, A. Kawka) for obtaining FEROS data. Parts of the observations obtained with the MPG 2.2m telescope were supported by the Ministry of Education, Youth and Sports project - LG14013 (Tycho Brahe: Supporting Ground-based Astronomical Observations). We also thank R. Venero for fruitful discussions. A.F.T., L.S.C. and M.L.A. acknowledge financial support from the Universidad Nacional de La Plata (Programa de Incentivos G11/137 and Proyecto Promocional de Investigaci\'on y Desarrollo G003) and the CONICET (PIP 0177), Argentina. M.K. acknowledges financial support from GA\v{C}R (grant number 17-02337S) and from the European Union European Regional Development Fund, project Benefits for Estonian Society from Space Research and Application (KOMEET, 2014\,-\,2020.\,4.\,01.\,16\,-\,0029). GM acknowledges support from CONICYT, Programa de Astronom\'ia/PCI, FONDO ALMA 2014, Proyecto No 31140024. Financial support for the international cooperation of the Czech Republic and Argentina (AVCR-CONICET/14/003) is acknowledged. The Astronomical Institute Ond\v{r}ejov is supported by the project RVO:67985815. This work was also supported by the institutional research funding IUT40-1 of the Estonian Ministry of Education and Research.
\end{acknowledgements}

%______________________________________________________________________________________
% FIGURE 21-----------------------------
% -------------------------------------
%______________________________________________________________________________________
%______________________________________________________________________________________

\bibliographystyle{./aa}
\bibliography{./biblio-s35}

\begin{thebibliography}{69}
\expandafter\ifx\csname natexlab\endcsname\relax\def\natexlab#1{#1}\fi

\bibitem[{{Aret} {et~al.}(2017){Aret}, {Kolka}, {Kraus}, \&
  {Maravelias}}]{are17}
{Aret}, A., {Kolka}, I., {Kraus}, M., \& {Maravelias}, G. 2017, in Astronomical
  Society of the Pacific Conference Series, Vol. 508, The B[e] Phenomenon:
  Forty Years of Studies, ed. A.~{Miroshnichenko}, S.~{Zharikov}, D.~{Kor{\v
  c}{\'a}kov{\'a}}, \& M.~{Wolf}, 239

\bibitem[{{Aret} {et~al.}(2016){Aret}, {Kraus}, \& {{\v S}lechta}}]{are16a}
{Aret}, A., {Kraus}, M., \& {{\v S}lechta}, M. 2016, \mnras, 456, 1424

\bibitem[{{Bartlett} {et~al.}(2013){Bartlett}, {Clark}, {Coe}, {Garcia}, \&
  {Uttley}}]{bar13}
{Bartlett}, E.~S., {Clark}, J.~S., {Coe}, M.~J., {Garcia}, M.~R., \& {Uttley},
  P. 2013, \mnras, 429, 1213

\bibitem[{{Bitsakis} {et~al.}(2017){Bitsakis}, {Bonfini},
  {Gonz{\'a}lez-L{\'o}pezlira}, {Ram{\'{\i}}rez-Siordia}, {Bruzual}, {Charlot},
  {Maravelias}, \& {Zaritsky}}]{bit17}
{Bitsakis}, T., {Bonfini}, P., {Gonz{\'a}lez-L{\'o}pezlira}, R.~A., {et~al.}
  2017, \apj, 845, 56

\bibitem[{{Bonanos} {et~al.}(2009){Bonanos}, {Massa}, {Sewilo}, {Lennon},
  {Panagia}, {Smith}, {Meixner}, {Babler}, {Bracker}, {Meade}, {Gordon},
  {Hora}, {Indebetouw}, \& {Whitney}}]{bon09}
{Bonanos}, A.~Z., {Massa}, D.~L., {Sewilo}, M., {et~al.} 2009, AJ, 138, 1003

\bibitem[{{Chita} {et~al.}(2008){Chita}, {Langer}, {van Marle},
  {Garc{\'{\i}}a-Segura}, \& {Heger}}]{chi08}
{Chita}, S.~M., {Langer}, N., {van Marle}, A.~J., {Garc{\'{\i}}a-Segura}, G.,
  \& {Heger}, A. 2008, \aap, 488, L37

\bibitem[{{Cidale} {et~al.}(2012){Cidale}, {Borges Fernandes}, {Andruchow},
  {Arias}, {Kraus}, {Chesneau}, {Kanaan}, {Cur{\'e}}, {de Wit}, \&
  {Muratore}}]{cid12}
{Cidale}, L.~S., {Borges Fernandes}, M., {Andruchow}, I., {et~al.} 2012, \aap,
  548, A72

\bibitem[{{Clark} {et~al.}(2013{\natexlab{a}}){Clark}, {Bartlett}, {Coe},
  {Dorda}, {Haberl}, {Lamb}, {Negueruela}, \& {Udalski}}]{cla13a}
{Clark}, J.~S., {Bartlett}, E.~S., {Coe}, M.~J., {et~al.} 2013{\natexlab{a}},
  \aap, 560, A10

\bibitem[{{Clark} {et~al.}(2000){Clark}, {Miroshnichenko}, {Larionov}, {Lyuty},
  {Hynes}, {Pooley}, {Coe}, {McCollough}, {Dieters}, {Efimov}, {Fabregat},
  {Goranskii}, {Haswell}, {Metlova}, {Robinson}, {Roche}, {Shenavrin}, \&
  {Welsh}}]{cla00}
{Clark}, J.~S., {Miroshnichenko}, A.~S., {Larionov}, V.~M., {et~al.} 2000,
  \aap, 356, 50

\bibitem[{{Clark} {et~al.}(2013{\natexlab{b}}){Clark}, {Ritchie}, \&
  {Negueruela}}]{cla13b}
{Clark}, J.~S., {Ritchie}, B.~W., \& {Negueruela}, I. 2013{\natexlab{b}}, \aap,
  560, A11

\bibitem[{{Conti}(1997)}]{con97}
{Conti}, P.~S. 1997, in Astronomical Society of the Pacific Conference Series,
  Vol. 120, Luminous Blue Variables: Massive Stars in Transition, ed. A.~{Nota}
  \& H.~{Lamers}, 161

\bibitem[{{Crowther} {et~al.}(2006){Crowther}, {Lennon}, {Walborn}, \&
  {Smartt}}]{cro06}
{Crowther}, P.~A., {Lennon}, D.~J., {Walborn}, N.~R., \& {Smartt}, S.~J. 2006,
  ArXiv Astrophysics e-prints

\bibitem[{{Danchi} {et~al.}(2001){Danchi}, {Tuthill}, \& {Monnier}}]{dan01}
{Danchi}, W.~C., {Tuthill}, P.~G., \& {Monnier}, J.~D. 2001, \apj, 562, 440

\bibitem[{{Davies} {et~al.}(2007){Davies}, {Oudmaijer}, \& {Sahu}}]{dav07}
{Davies}, B., {Oudmaijer}, R.~D., \& {Sahu}, K.~C. 2007, \apj, 671, 2059

\bibitem[{{de Jager}(1998)}]{dej98}
{de Jager}, C. 1998, \aapr, 8, 145

\bibitem[{{Domiciano de Souza} {et~al.}(2007){Domiciano de Souza}, {Driebe},
  {Chesneau}, {Hofmann}, {Kraus}, {Miroshnichenko}, {Ohnaka}, {Petrov},
  {Preisbisch}, {Stee}, {Weigelt}, {Lisi}, {Malbet}, \& {Richichi}}]{dom07}
{Domiciano de Souza}, A., {Driebe}, T., {Chesneau}, O., {et~al.} 2007, A\&A,
  464, 81

\bibitem[{{Dunstall} {et~al.}(2012){Dunstall}, {Fraser}, {Clark}, {Crowther},
  {Dufton}, {Evans}, {Lennon}, {Soszy{\'n}ski}, {Taylor}, \& {Vink}}]{dun12}
{Dunstall}, P.~R., {Fraser}, M., {Clark}, J.~S., {et~al.} 2012, \aap, 542, A50

\bibitem[{{Ekstr{\"o}m} {et~al.}(2012){Ekstr{\"o}m}, {Georgy}, {Eggenberger},
  {Meynet}, {Mowlavi}, {Wyttenbach}, {Granada}, {Decressin}, {Hirschi},
  {Frischknecht}, {Charbonnel}, \& {Maeder}}]{eks12}
{Ekstr{\"o}m}, S., {Georgy}, C., {Eggenberger}, P., {et~al.} 2012, \aap, 537,
  A146

\bibitem[{{Fenech} {et~al.}(2017){Fenech}, {Clark}, {Prinja}, {Morford},
  {Dougherty}, \& {Blomme}}]{fen17}
{Fenech}, D.~M., {Clark}, J.~S., {Prinja}, R.~K., {et~al.} 2017, \mnras, 464,
  L75

\bibitem[{{Fukagawa} {et~al.}(2004){Fukagawa}, {Hayashi}, {Tamura}, {Itoh},
  {Hayashi}, {Oasa}, {Takeuchi}, {Morino}, {Murakawa}, {Oya}, {Yamashita},
  {Suto}, {Mayama}, {Naoi}, {Ishii}, {Pyo}, {Nishikawa}, {Takato}, {Usuda},
  {Ando}, {Iye}, {Miyama}, \& {Kaifu}}]{fuk04}
{Fukagawa}, M., {Hayashi}, M., {Tamura}, M., {et~al.} 2004, \apjl, 605, L53

\bibitem[{{Glatt} {et~al.}(2010){Glatt}, {Grebel}, \& {Koch}}]{gla10}
{Glatt}, K., {Grebel}, E.~K., \& {Koch}, A. 2010, \aap, 517, A50

\bibitem[{{Groh} \& {Vink}(2011)}]{gro11}
{Groh}, J.~H. \& {Vink}, J.~S. 2011, \aap, 531, L10

\bibitem[{{Gummersbach} {et~al.}(1995){Gummersbach}, {Zickgraf}, \&
  {Wolf}}]{gum95}
{Gummersbach}, C.~A., {Zickgraf}, F.-J., \& {Wolf}, B. 1995, A\&A, 302, 409

\bibitem[{{Henize}(1956)}]{hen56}
{Henize}, K.~G. 1956, ApJS, 2, 315

\bibitem[{{Jaschek} {et~al.}(1993){Jaschek}, {Jaschek}, \& {Andrillat}}]{jas93}
{Jaschek}, M., {Jaschek}, C., \& {Andrillat}, Y. 1993, \aaps, 97, 781

\bibitem[{{Jorgenson} {et~al.}(2000){Jorgenson}, {Kogan}, \&
  {Strelnitski}}]{jor00}
{Jorgenson}, R.~A., {Kogan}, L.~R., \& {Strelnitski}, V. 2000, \aj, 119, 3060

\bibitem[{{Kaler} \& {Jacoby}(1991)}]{kal91}
{Kaler}, J.~B. \& {Jacoby}, G.~H. 1991, \apj, 382, 134

\bibitem[{{Kim} \& {Taam}(2012)}]{kim12}
{Kim}, H. \& {Taam}, R.~E. 2012, \apj, 759, 59

\bibitem[{{Ko} {et~al.}(2013){Ko}, {Lee}, \& {Lim}}]{ko13}
{Ko}, Y., {Lee}, M.~G., \& {Lim}, S. 2013, \apj, 777, 82

\bibitem[{{Kraus}(2009)}]{kra09}
{Kraus}, M. 2009, \aap, 494, 253

\bibitem[{{Kraus} {et~al.}(2016){Kraus}, {Cidale}, {Arias}, {Maravelias},
  {Nickeler}, {Torres}, {Borges Fernandes}, {Aret}, {Cur{\'e}},
  {Vallverd{\'u}}, \& {Barb{\'a}}}]{kra16}
{Kraus}, M., {Cidale}, L.~S., {Arias}, M.~L., {et~al.} 2016, \aap, 593, A112

\bibitem[{{Kraus} {et~al.}(2014){Kraus}, {Cidale}, {Arias}, {Oksala}, \&
  {Borges Fernandes}}]{Kra14}
{Kraus}, M., {Cidale}, L.~S., {Arias}, M.~L., {Oksala}, M.~E., \& {Borges
  Fernandes}, M. 2014, \apjl, 780, L10

\bibitem[{{Kraus} {et~al.}(2000){Kraus}, {Kr{\"u}gel}, {Thum}, \&
  {Geballe}}]{kra00}
{Kraus}, M., {Kr{\"u}gel}, E., {Thum}, C., \& {Geballe}, T.~R. 2000, \aap, 362,
  158

\bibitem[{{Kraus} {et~al.}(2017){Kraus}, {Liimets}, {Cappa}, {Cidale},
  {Nickeler}, {Duronea}, {Arias}, {Gunawan}, {Oksala}, {Borges Fernandes},
  {Maravelias}, {Cur{\'e}}, \& {Santander-Garc{\'{\i}}a}}]{kra17}
{Kraus}, M., {Liimets}, T., {Cappa}, C.~E., {et~al.} 2017, \aj, 154, 186

\bibitem[{{Lamers} {et~al.}(1998){Lamers}, {Zickgraf}, {de Winter}, {Houziaux},
  \& {Zorec}}]{lam98}
{Lamers}, H.~J.~G.~L.~M., {Zickgraf}, F.-J., {de Winter}, D., {Houziaux}, L.,
  \& {Zorec}, J. 1998, \aap, 340, 117

\bibitem[{{Levato} {et~al.}(2014){Levato}, {Miroshnichenko}, \&
  {Saffe}}]{Lev14}
{Levato}, H., {Miroshnichenko}, A.~S., \& {Saffe}, C. 2014, \aap, 568, A28

\bibitem[{{Liermann} {et~al.}(2010){Liermann}, {Kraus}, {Schnurr}, \&
  {Fernandes}}]{lie10}
{Liermann}, A., {Kraus}, M., {Schnurr}, O., \& {Fernandes}, M.~B. 2010, \mnras,
  408, L6

\bibitem[{{Lortet} \& {Testor}(1988)}]{lor88}
{Lortet}, M.-C. \& {Testor}, G. 1988, \aap, 194, 11

\bibitem[{{Magalh{\~a}es} {et~al.}(2006){Magalh{\~a}es}, {Melgarejo},
  {Pereyra}, \& {Carciofi}}]{mag06}
{Magalh{\~a}es}, A.~M., {Melgarejo}, R., {Pereyra}, A., \& {Carciofi}, A.~C.
  2006, in Astronomical Society of the Pacific Conference Series, Vol. 355,
  Stars with the B[e] Phenomenon, ed. M.~{Kraus} \& A.~S. {Miroshnichenko}, 147

\bibitem[{{Magalhaes}(1992)}]{mag92}
{Magalhaes}, A.~M. 1992, ApJ, 398, 286

\bibitem[{{Maravelias} {et~al.}(2017){Maravelias}, {Kraus}, {Aret}, {Cidale},
  {Arias}, \& {Borges Fernandes}}]{mar17}
{Maravelias}, G., {Kraus}, M., {Aret}, A., {et~al.} 2017, in Astronomical
  Society of the Pacific Conference Series, Vol. 508, The B[e] Phenomenon:
  Forty Years of Studies, ed. A.~{Miroshnichenko}, S.~{Zharikov}, D.~{Kor{\v
  c}{\'a}kov{\'a}}, \& M.~{Wolf}, 213

\bibitem[{{Maravelias} {et~al.}(2014){Maravelias}, {Zezas}, {Antoniou}, \&
  {Hatzidimitriou}}]{mar14}
{Maravelias}, G., {Zezas}, A., {Antoniou}, V., \& {Hatzidimitriou}, D. 2014,
  \mnras, 438, 2005

\bibitem[{{Markova} \& {Puls}(2008)}]{mar08}
{Markova}, N. \& {Puls}, J. 2008, \aap, 478, 823

\bibitem[{{Massey} {et~al.}(2014){Massey}, {Neugent}, {Morrell}, \&
  {Hillier}}]{mas14}
{Massey}, P., {Neugent}, K.~F., {Morrell}, N., \& {Hillier}, D.~J. 2014, \apj,
  788, 83

\bibitem[{{Mathew} {et~al.}(2012){Mathew}, {Banerjee}, {Subramaniam}, \&
  {Ashok}}]{mat12}
{Mathew}, B., {Banerjee}, D.~P.~K., {Subramaniam}, A., \& {Ashok}, N.~M. 2012,
  \apj, 753, 13

\bibitem[{{Melgarejo} {et~al.}(2001){Melgarejo}, {Magalh{\~a}es}, {Carciofi},
  \& {Rodrigues}}]{mel01}
{Melgarejo}, R., {Magalh{\~a}es}, A.~M., {Carciofi}, A.~C., \& {Rodrigues},
  C.~V. 2001, A\&A, 377, 581

\bibitem[{{Millour} {et~al.}(2011){Millour}, {Meilland}, {Chesneau}, {Stee},
  {Kanaan}, {Petrov}, {Mourard}, \& {Kraus}}]{mil11}
{Millour}, F., {Meilland}, A., {Chesneau}, O., {et~al.} 2011, \aap, 526, A107

\bibitem[{{Oksala} {et~al.}(2013){Oksala}, {Kraus}, {Cidale}, {Muratore}, \&
  {Borges Fernandes}}]{oks13}
{Oksala}, M.~E., {Kraus}, M., {Cidale}, L.~S., {Muratore}, M.~F., \& {Borges
  Fernandes}, M. 2013, \aap, 558, A17

\bibitem[{{Oudmaijer} \& {Drew}(1999)}]{oud99}
{Oudmaijer}, R.~D. \& {Drew}, J.~E. 1999, \mnras, 305, 166

\bibitem[{{Penny} \& {Gies}(2009)}]{pen09}
{Penny}, L.~R. \& {Gies}, D.~R. 2009, ApJ, 700, 844

\bibitem[{{P{\'e}rez} {et~al.}(2016){P{\'e}rez}, {Carpenter}, {Andrews},
  {Ricci}, {Isella}, {Linz}, {Sargent}, {Wilner}, {Henning}, {Deller},
  {Chandler}, {Dullemond}, {Lazio}, {Menten}, {Corder}, {Storm}, {Testi},
  {Tazzari}, {Kwon}, {Calvet}, {Greaves}, {Harris}, \& {Mundy}}]{per16}
{P{\'e}rez}, L.~M., {Carpenter}, J.~M., {Andrews}, S.~M., {et~al.} 2016,
  Science, 353, 1519

\bibitem[{{Pojmanski}(2002)}]{poj02}
{Pojmanski}, G. 2002, \actaa, 52, 397

\bibitem[{{Reid}(2014)}]{rei14}
{Reid}, W.~A. 2014, \mnras, 438, 2642

\bibitem[{{Sallum} {et~al.}(2017){Sallum}, {Eisner}, {Hinz}, {Sheehan},
  {Skemer}, {Tuthill}, \& {Young}}]{sal17}
{Sallum}, S., {Eisner}, J.~A., {Hinz}, P.~M., {et~al.} 2017, \apj, 844, 22

\bibitem[{{Schwarzenberg-Czerny}(1989)}]{sch89}
{Schwarzenberg-Czerny}, A. 1989, \mnras, 241, 153

\bibitem[{{Schwarzenberg-Czerny}(1996)}]{sch96}
{Schwarzenberg-Czerny}, A. 1996, \apjl, 460, L107

\bibitem[{{Schwarzenberg-Czerny} \& {Beaulieu}(2006)}]{sch06}
{Schwarzenberg-Czerny}, A. \& {Beaulieu}, J.-P. 2006, \mnras, 365, 165

\bibitem[{{Shu}(2016)}]{shu16}
{Shu}, F.~H. 2016, \araa, 54, 667

\bibitem[{{Smith} {et~al.}(2011){Smith}, {Gehrz}, {Campbell}, {Kassis}, {Le
  Mignant}, {Kuluhiwa}, \& {Filippenko}}]{smi11}
{Smith}, N., {Gehrz}, R.~D., {Campbell}, R., {et~al.} 2011, \mnras, 418, 1959

\bibitem[{{Stahl}(2001)}]{sta01}
{Stahl}, O. 2001, in Astronomical Society of the Pacific Conference Series,
  Vol. 242, Eta Carinae and Other Mysterious Stars: The Hidden Opportunities of
  Emission Spectroscopy, ed. T.~R. {Gull}, S.~{Johannson}, \& K.~{Davidson},
  163

\bibitem[{{Tafoya} {et~al.}(2004){Tafoya}, {G{\'o}mez}, \&
  {Rodr{\'{\i}}guez}}]{taf04}
{Tafoya}, D., {G{\'o}mez}, Y., \& {Rodr{\'{\i}}guez}, L.~F. 2004, \apj, 610,
  827

\bibitem[{{Torres} {et~al.}(2012){Torres}, {Kraus}, {Cidale}, {Barb{\'a}},
  {Borges Fernandes}, \& {Brandi}}]{tor12}
{Torres}, A.~F., {Kraus}, M., {Cidale}, L.~S., {et~al.} 2012, \mnras, 427, L80

\bibitem[{{van Genderen} \& {Sterken}(2002)}]{gen02}
{van Genderen}, A.~M. \& {Sterken}, C. 2002, \aap, 386, 926

\bibitem[{{van Loon} {et~al.}(2005){van Loon}, {Marshall}, \&
  {Zijlstra}}]{loo05}
{van Loon}, J.~T., {Marshall}, J.~R., \& {Zijlstra}, A.~A. 2005, \aap, 442, 597

\bibitem[{{Vink}(2000)}]{vin00}
{Vink}, J.~S. 2000, PhD thesis, Universiteit Utrecht, \email{jvink@phys.uu.nl}

\bibitem[{{Wright} {et~al.}(2010){Wright}, {Eisenhardt}, {Mainzer}, {Ressler},
  {Cutri}, {Jarrett}, {Kirkpatrick}, {Padgett}, {McMillan}, {Skrutskie},
  {Stanford}, {Cohen}, {Walker}, {Mather}, {Leisawitz}, {Gautier}, {McLean},
  {Benford}, {Lonsdale}, {Blain}, {Mendez}, {Irace}, {Duval}, {Liu}, {Royer},
  {Heinrichsen}, {Howard}, {Shannon}, {Kendall}, {Walsh}, {Larsen}, {Cardon},
  {Schick}, {Schwalm}, {Abid}, {Fabinsky}, {Naes}, \& {Tsai}}]{wri10}
{Wright}, E.~L., {Eisenhardt}, P.~R.~M., {Mainzer}, A.~K., {et~al.} 2010, \aj,
  140, 1868

\bibitem[{{Yudin}(1996)}]{yud96}
{Yudin}, R.~V. 1996, \aap, 312, 234

\bibitem[{{Zickgraf} {et~al.}(1986){Zickgraf}, {Wolf}, {Leitherer},
  {Appenzeller}, \& {Stahl}}]{zic86}
{Zickgraf}, F.-J., {Wolf}, B., {Leitherer}, C., {Appenzeller}, I., \& {Stahl},
  O. 1986, \aap, 163, 119

\bibitem[{{Zickgraf} {et~al.}(1989){Zickgraf}, {Wolf}, {Stahl}, \&
  {Humphreys}}]{zic89}
{Zickgraf}, F.-J., {Wolf}, B., {Stahl}, O., \& {Humphreys}, R.~M. 1989, A\&A,
  220, 206

\end{thebibliography}

\end{document}